\documentclass[12pt]{iopart}

\usepackage{iopams}  
\usepackage{graphicx}
\usepackage{xcolor}
\newcommand{\bq}{\begin{equation}}
\newcommand{\eq}{\end{equation}}
\newcommand{\bqn}{\begin{eqnarray}}
\newcommand{\eqn}{\end{eqnarray}}
\newcommand{\nb}{\nonumber}
\newcommand{\lb}{\label}

\begin{document}
\title{{Vacuum} solutions in the Einstein-Aether Theory}

\author{M. Campista $^{1,2}$, R. Chan $^{1}$, M. F. A. da Silva $^{3}$, O. Goldoni $^{3}$,\\ V. H. Satheeshkumar$^{1,3,4}$ and Jaime F. Villas da Rocha $^5$} 

\address{$^{1}$Coordena\c{c}\~ao de Astronomia e Astrof\'{\i}sica, 
	Observat\'orio Nacional (ON), Rua General Jos\'e Cristino, 77, S\~ao Crist\'ov\~ao, CEP 20921-400, 	Rio de Janeiro, RJ, Brazil.\\
	
	$^{2}$Departamento de F\'{\i}sica Te\'orica,
	Instituto de F\'{\i}sica, Universidade Federal do Rio de Janeiro (UFRJ), Rio de Janeiro, Brazil.\\
	
	$^{3}$Departamento de F\'{\i}sica Te\'orica,
	Instituto de F\'{\i}sica, Universidade do Estado do Rio de Janeiro (UERJ),
	Rua S\~ao Francisco Xavier 524, Maracan\~a,
	CEP 20550-900, Rio de Janeiro, RJ, Brazil.\\
	
	$^{4}$Centro Brasileiro de Pesquisas F\'{i}sicas (CBPF), 
	Rua Dr. Xavier Sigaud, 150, Urca, CEP 22290-180, Rio de Janeiro, RJ, Brazil.\\ 
	
	$^{5}$Universidade Federal do Estado do Rio de Janeiro (UNIRIO),
	Instituto de Bioci\^encias,
	Departamento de Ci\^encias Naturais, Avenida Pasteur 458, Urca,
	CEP 22290-240, Rio de Janeiro, RJ, Brazil.}
	
\ead{campista@on.br, chan@on.br, mfasnic@gmail.com, otaviosama@gmail.com, vhsatheeshkumar@gmail.com, jfvroch@gmail.com}

\begin{abstract}
	
The Einstein-Aether (EA) theory belongs to a class of modified gravity theories characterized by the introduction of a time-like unit vector field, called aether. In this scenario, a preferred frame arises as a natural consequence of a broken Lorentz invariance. In the present work we have obtained and  analyzed   some  exact solutions allowed by this theory for two particular cases {of perfect fluid, both with Friedmann-Lema{\^{\i}}tre-Robertson-Walker (FLRW) symmetry}: (i) a fluid with constant energy density 
{($p=-\rho_0$)}, and (ii) a fluid with {zero} energy density 
{($\rho_0=0$)}
{, corresponding to the vacuum solution with and without cosmological constant {($\Lambda$)}, respectively. 
Our solutions show
that the EA and GR theories do not differentiate each other only by the
coupling constants. This difference is clearly shown because of the existence
of singularities that there are not in GR theory.} {This characteristic appears in the solutions with
$p=-\rho_0$ as well as with $\rho_0=0$, where this last one depends only on the aether field.} 
{Besides, we consider the term of the EA theory in the Raychaudhuri equation and discuss the meaning of the strong energy condition in this scenario and found that this depends on aether field.} 
The solutions admit an expanding or contracting system. A bounce, a singular, a constant and an accelerated expansion solution{s} were also obtained, exhibiting the richness of the EA theory from the dynamic point of view of a collapsing system or of a cosmological model. The analysis of energy conditions, {considering an effective fluid} shows that the term of the  aether contributes significantly for the accelerated expansion of the system for the case in which the energy density is constant. On the other hand, for the vacuum case {($\rho_0=0$)}, the energy conditions are all satisfied for the aether fluid. 
\end{abstract}


\maketitle

\section{Introduction}

A  generally  covariant  theory  in  which  local  Lorentz  Invariance (LI) is  broken  by  a  dynamical  unit timelike vector field $u^a$ often referred to as aether was newly introduced in 2001 to study the preferred frame effects in gravitation and cosmology \cite{Jacobson:2000xp}. The presence of a preferred frame defined by the timelike unit vector field or aether field breaks LI, which makes Einstein-Aether (EA) theory a low energy effective theory.  In fact, the action of EA theory is the most general generally covariant functional of the spacetime metric $g_{ab}$ and aether
field $u^a$ involving no more than two derivatives. 

The parameters of theory have been severely constrained using many observational/experimental tests such as the primordial nucleosynthesis \cite{Carroll:2004ai}, ultra-high energy cosmic rays \cite{Elliott:2005va}, the solar system tests \cite{Eling:2003rd, GrEAsser:2005bg}, binary pulsars \cite{Foster:2007gr, Yagi:2013ava}, and more recently gravitational waves \cite{Gong:2018cgj, Oost:2018tcv}.  These results confirm that there exists a family of EA theories with `small-enough' couplings that passes all current observational tests \cite{Foster:2005dk}.
There is an extensive literature about the EA field equation solutions \cite{Barrow2011, Bhattacharjee}.

Carroll and Lim \cite{Carroll:2004ai} presented the field equations for the EA theory considering the Lorentz-violating timelike vector field with fixed norm. They used the metric of FLRW since they were interested in cosmological solutions. 
{  One of their conclusions is that the introduction of the vector field changes the Newtonian coupling constant value and contributes to the decrease in the rate of expansion of the system, although the solutions for the field equations have not been obtained. We note that,  although the authors have presented the usual cosmological field equations of the standard model, they have based their conclusions only on one of them. However,  one should expect that the functions $ H (t)$ and $a (t)$  may not behave as in GR theory. Therefore, their conclusion ``{\it the net effect of the vector field is to decrease the rate of expansion of the universe}'' could only be verified, or not, with the consideration of the other two field equations.} {Other authors remark that this conclusion it is true only for small spatial curvature \cite{Jacobson08, Mattingly2001}}.
In this work, motivated by Carroll and Lim, we seek to obtain exact solutions for the field equations presented by them, but using the same formalism proposed by Garfinkle et al. \cite {Garfinkle}. Our purpose is to investigate the physical and geometric properties of the solutions obtained to better identify the possible differences between the two theories, EA and GR.

The paper is organized as follows.  The Section $2$ presents the EA field equations.
The Section $3$, describes the possibles solutions of the field equations. In Section $4$, the dynamics of these solutions are analyzed while in the Section $5$, are analyzed the energy condition.
Finally, Section $6$, presents our conclusions.

\section{Field Equations in the EA theory }

The general action of the EA theory, in a background where the  metric signature is $({-}{+}{+}{+})$ and the units are chosen so that the speed of light defined by the metric $g_{ab}$ is unity, is given by  

\bq S = \int \sqrt{-g}~(L_{\rm Einstein}+L_{\rm aether}+L_{\rm matter}) d^{4}x,
 \label{action} 
\eq
where, the first term is the usual Einstein-Hilbert Lagrangian, defined by $R$, the Ricci scalar, and $G_N$, the Newtonian gravitational constant, {as}
\bq L_{\rm Einstein} =  \frac{1}{16\pi G} R. \eq
{The} second term, the aether Lagrangian {is given by}
\bq L_{\rm aether} =  \frac{1}{16\pi G} [-K^{ab}{}_{mn} \nabla_a u^m
\nabla_b u^n +
\lambda(g_{ab}u^a u^b + 1)],
\lb{LEAG}
\eq
where the tensor ${K^{ab}}_{mn}$ is defined as
\bq {{K^{ab}}_{mn}} = c_1 g^{ab}g_{mn}+c_2\delta^{a}_{m} \delta^{b}_{n}
+c_3\delta^{a}_{n}\delta^{b}_{m}-c_4u^a u^b g_{mn},
\lb{Kab}
\eq
being the $c_i$ dimensionless coupling constants, and $\lambda$
a Lagrange multiplier enforcing the unit timelike constraint on the aether, and 
\bq
\delta^a_m \delta^b_n =g^{a\alpha}g_{\alpha m} g^{b\beta}g_{\beta n}.
\eq

Finally, the last term, $L_{\rm matter}$ is the matter Lagrangian, which depends on the metric tensor and the matter field.

 In the weak-field, slow-motion limit EA theory reduces to Newtonian gravity with a value of  Newton's constant $G_{\rm N}$ related to the parameter $G$ in the action (\ref{action})  by  { \cite{Garfinkle}},

 \bq
 G = G_N\left(1-\frac{c_1+c_4}{2}\right).
 \lb{Ge}
 \eq
 
Note that if $c_1=-c_4$ the EA coupling constant $G$ becomes the Newtonian coupling constant $G_N$, without {necessarily} imposing $c_1=c_4=0$.

The field equations are obtained by extremizing the action with respect to independent  variables of the system. The variation with respect to the Lagrange multiplier $\lambda$ imposes the condition that $u^a$ is a unit timelike vector, thus 
\bq
g_{ab}u^a u^b = -1,
\label{LagMul}
\eq
while the variation of the action with respect $u^a$, leads to \cite{Garfinkle}
\bq
 \nabla_a J^a_b + c_4 a_a \nabla_b u^a + \lambda u_b = 0,
\eq
where,
\bq
J^a_m=K^{ab}_{mn} \nabla_b u^n,
\eq
and
\bq
a_a=u^b \nabla_b u_a.
\eq
The variation of the action with respect to the metric $g_{mn}$ gives the dynamical equations,
\bq
G^{Einstein}_{ab} = T^{aether}_{ab} +8 \pi G  T^{matter}_{ab},
\label{EA}
\eq
where 
\bqn
G^{Einstein}_{ab} &=& R_{ab} - \frac{1}{2} g_{ab} R, \nb \\
T^{aether}_{ab}&=& \nabla_c [ J^c\;_{(a} u_{b)} + u^c J_{(ab)} - J_{(a} \;^c u_{b)}] - \frac{1}{2} g_{ab} J^c_d \nabla_c u^d+ \lambda u_a u_b  \nb \\
& & + c_1 [\nabla_a u_c \nabla_b u^c - \nabla^c u_a \nabla_c u_b] + c_4 a_a a_b, \nb \\
T^{matter}_{ab} &=&  \frac{- 2}{\sqrt{-g}} \frac{\delta \left( \sqrt{-g} L_{matter} \right)}{\delta g_{ab}}.
\label{fieldeqs}
\eqn

Sticking to the convention, we have considered aether on the matter-side of the field equations. It might as well be on the geometry-side { we have the
effect of the aether vector field} which is reflected by the fact that the coupling constant of the EA theory is different from that of GR. So, it is unfortunate if one gets an impression that EA theory is simply GR coupled to vector field because of equation (\ref{fieldeqs}).  

{In a more general situation, t}he Lagrangian of GR theory is recovered, if and only if, the coupling constants are identically null, e.g., $c_1=c_2=c_3=c_4=0$, 
{considering} the equations 
{(\ref{Kab}) and (\ref{LagMul})}.

Aiming to know what kind of solutions the EA theory admits {with spacial constant} curvature, we will assume a FLRW metric
\bq
ds^2= -dt^2+a(t)^2\left[\frac{dr^2}{1-kr^2} +r^2 d\theta^2 +r^2 \sin^2 \theta d\phi^2\right],
\lb{ds2}
\eq
where, $a(t)$ is the scale factor {and} $k$ is a constant representing the Gaussian curvature of the space at a given time. 

{The choice of a FLRW metric is natural, since in  the framework of GR theory it provides} the basis for most problems in cosmology and can also be useful in studies of gravitational collapse
(since the Oppenheimer-Snyder's landmark work \cite{OS_1939, Kolassis1988, Chan1989, Sharif2010} { and should be interesting to explore it in the EA theory}.

In accordance with equation (\ref{LagMul}), the aether field is assumed unitary, timelike and constant, chosen as
\bq
u^a=(1,0,0,0).
\eq

Assuming (\ref{ds2}), we compute the different terms in the field equations equation (\ref{fieldeqs}). Firstly, the usual geometric part of GR is represented by, 
\bq
G^{Einstein}_{tt}=\frac{3}{a^2} \left[k+\dot a^2\right],
\lb{Gtt}
\eq
\bq
G^{Einstein}_{rr}=-\frac{1}{1-k r^2}\left[k+\dot a^2+2 a \ddot a\right],
\lb{Grr}
\eq
\bq
G^{Einstein}_{\theta\theta}=-(k r^2+r^2 \dot a^2+2 r^2 a \ddot a),
\eq
\bq
G^{Einstein}_{\phi\phi}=G^{Einstein}_{\theta\theta} \sin^2 \theta.
\eq
The contribution of the aether field to the stress-energy tensor is represented by,
\bq
T^{aether}_{tt}=-\frac{3\beta}{2} \frac{\dot a^2}{a^2},
\lb{Ttt}
\eq
\bq
T^{aether}_{rr}=\frac{\beta}{2(1-k r^2)} \left[\dot a^2+2 a \ddot a\right],
\lb{Trr}
\eq
\bq
T^{aether}_{\theta\theta}=\frac{r^2 \beta}{2}  \left[\dot a^2+2 a \ddot a\right],
\eq
\bq
T^{aether}_{\phi\phi}=T^{aether}_{\theta\theta} \sin^2 \theta.
\lb{Tphiphi}
\eq
Here, the constant $\beta$ is defined as
\bq
\beta=c_1+3 c_2+c_3.
\lb{beta}
\eq

{It is {easy} to see that the Lagrangian reduces to
\bq
L_{\rm aether} = -\frac{3 \beta}{16\pi G}
\left({\frac {\dot a}{a}}\right)^{2}.
\lb{LEA}
\eq
}

{Note that for
the choices made here the Lagrangian and the equations (\ref{Ttt})-(\ref{Tphiphi}) {vanish} for $\beta=0$,
independently the values of $c_1$, $c_2$ and $c_3$.}
{Moreover, we call attention that even if $\beta=0$ and $c_1=-c_4$, although the field equations coincide with that from GR theory, the preferential frame still exist.} 

It can be prove, by equations (\ref{Gtt})-(\ref{Tphiphi}), that we can not have an anisotropic pressure fluid, corroborating the representation of matter as a perfect fluid 
\bq
T^{matter}_{ab}=(\rho+p) v_a v_b + p g_{ab},
\lb{TEA}
\eq
where $v^a$ is a four-velocity of the matter fluid. For sake of simplicity, we will consider $v^a$ in a comoving frame with aether fluid, reason for what it can be represented by  $v^a=\delta^a_t$, a time like unitary vector.

{
Combining equations (\ref{Gtt}), (\ref{Grr}), (\ref{Ttt}) and (\ref{Trr}), we obtain the analogue of the Friedmann equations, that are
\bq
\frac{\ddot a}{a}=-\frac{8 \pi G}{3(\beta+2)}(\rho+3p),
\lb{Fried1}
\eq
\bq
\left(\frac{\dot a}{a}\right)^2=\frac{16 \pi G \rho}{3(\beta+2)}-\frac{2k}{a^2(\beta+2)}.
\lb{Fried2}
\eq
}

In addition, it is important to require that the divergence of the energy-momentum tensor be zero, {implying in the following  equation of continuity}
\begin{equation}
\label{DEMT}
\dot\rho+ 3\frac{\dot a}{a}(\rho+p)=0.
\end{equation}

In order to identify eventual singularities in the solutions, is very useful to calculate the Kretschmann scalar invariant K. For the metric (\ref{ds2}), it is  given by
\bq
K   = \frac{12}{a^4} \left[k^2+2 k \dot a^2+\dot a^4+\ddot a^2 a^2\right],
\lb{K}
\eq
where $a(t)$ will be the solutions of (\ref{Gtt})-(\ref{Tphiphi}) and (\ref{TEA}).

\section{Solutions of EA field equations}

In this section,  we present and analyze solutions of field equations (\ref{Gtt})-(\ref{Tphiphi}) for  following particular cases of (\ref{TEA})

\subsection{ {Vacuum solutions with $\Lambda>0$}}
{Vacuum solution with $\Lambda >0$ is equivalent to
an} isotropic and homogeneous fluid of pressure $p$ and constant density  $\rho=\rho_0$, {here} we obtain {a} set of exact solutions for $\dot a(t)\neq 0$. {The case $\dot a(t)= 0$} is not relevant since the aether field {contribution to the field equations} vanishes. Then, for $\dot a(t)\neq 0$, we have
{
\bq
a(t)=\frac{ \sqrt{6}}{6w (\beta+2)} \left[e^{- \alpha w(t-t_0)}-6\left(\frac{\beta}{2}+1\right)e^{\alpha w (t-t_0)}\right], \;\;\;k=-1,
\lb{Bt1}
\eq
\bq
a(t)=e^{-\alpha w (t-t_0)}, \;\;\;k=0,
\lb{Bt2}
\eq
\bq
a(t)=\frac{ \sqrt{6}}{6w (\beta+2)} \left[e^{- \alpha w (t-t_0)}+6\left(\frac{\beta}{2}+1\right)e^{\alpha w (t-t_0)}\right] ,\;\;\;k=1,
\lb{Bt3}
\eq
}
where $\alpha=\pm 1$ and 
\bq
w=4\sqrt{\frac{\pi G \rho_0}{3(\beta+2)}},
\lb{omega}
\eq
with $\beta+2>0$ in order to have $w$ real and not divergent. { The equation of continuity (\ref{DEMT}) imposes $p=-\rho_0$}, {as expected to the vacuum with $\Lambda>0$}, characterizing a dark energy fluid in GR theory.
{However in EA theory, as seen in equation (\ref{EA}), there is an additional term depending on $\beta$, which {would change} this interpretation. 
Thus, in Section 5, we interpret the role of this additional term. }
Note that there is an integration constant $t_0$ which must be chosen in a such way that we have the Sitter solutions at the limit of GR theory.

It is easy to see that equations (\ref{Bt1})-(\ref{Bt3}) reduce to the de Sitter in GR theory \cite{Stephani1990} since we assume that $\beta=0$ and $c_1=-c_4$ {\footnote{It is possible $c_1 \neq -c_4$, rescaling the cosmological constant.}}
\bq
t_0={\frac {\ln  \left( 6 \right) }{2\alpha\,{\it w_0}}},
\lb{t0}
\eq
{with the cosmological constant $\Lambda=3\,{{\it w_0}}^{2}$ and 
\bq
w_0=4\sqrt{\frac{\pi G_N \rho_0}{6}}.
\eq
}
{
Here we call attention to the fact that in general, in the literature \cite{Stephani1990, Rindler2006, Padmanabhan2010, d'Inverno1992}
it is assumed the de Sitter spacetime only when $\alpha=-1$, $a(t) \ge 0$ and $t \ge 0$  (expanding systems). However, the de Sitter equations do not forbid solutions with  $\alpha=1$ , $a(t) < 0$ or $t < 0$ (contracting systems), which can represent collapsing systems {\cite{OS_1939}-\cite{Sharif2010}}. 
}

{Thus, considering equation (\ref{t0}), the equations (\ref{Bt1})-(\ref{Bt3}) can be rewritten as
\bq
a(t)= \frac{1}{ {w}\left( \beta+2 \right) {6}^{\frac{1}{2}+\frac{1}{2}\,{\frac {w}{{\it w_0}}}}}\left[ {6}^{{
\frac {w}{{\it w_0}}}}{{\rm e}^{-\alpha\,wt}}- 6\left(\frac{\beta}{2}+1\right) {{\rm e}^{\alpha\,wt}} \right],k=-1,
\label{dS1}
\eq
\bq
a(t)=\sqrt {6}{{\rm e}^{-\alpha\,{\it w}\,t}},\;k=0,
\lb{dS2}
\eq
\bq
a(t)=\frac{1}{ {w}\left( \beta+2 \right) {6}^{\frac{1}{2}+\frac{1}{2}\,{\frac {w}{{\it w_0}}}}}\left[ {6}^{{
\frac {w}{{\it w_0}}}}{{\rm e}^{-\alpha\,wt}}+ 6\left(\frac{\beta}{2}+1\right) {{\rm e}^{\alpha\,wt}} \right],k=1.
\lb{dS3}
\eq
}

From now on, we will assume {$t_0=0$} for the sake of simplicity.

{By the computation of Kretschmann scalar for the solutions (\ref{Bt1})-(\ref{Bt3}), we} found that a unique divergent solution {occurs for} $k=-1$, and the {corresponding} Kretschmann scalar is
\bqn
K&=&24 {w}^{4}\, \left( 216+360\,\beta+270\,{\beta}^{2}+108\,{\beta}^{3}-
24\,{{\rm e}^{-2\,\alpha\,wt}}+1296\,{{\rm e}^{4\,\alpha\,wt}}\right. \nb\\
&&\left. -1296\,{
	{\rm e}^{2\,\alpha\,wt}}{\beta}^{2}+648\,{{\rm e}^{4\,\alpha\,wt}}{
	\beta}^{3}+1944\,{{\rm e}^{4\,\alpha\,wt}}{\beta}^{2}-432\,{{\rm e}^{2
		\,\alpha\,wt}}{\beta}^{3}-\right.\nb\\
&&\left. 54\,{{\rm e}^{2\,\alpha\,wt}}{\beta}^{4}+
2592\,{{\rm e}^{4\,\alpha\,wt}}\beta+81\,{{\rm e}^{4\,\alpha\,wt}}{
	\beta}^{4}-24\,{{\rm e}^{-2\,\alpha\,wt}}\beta-\right.\nb \\
&&\left. 6\,{{\rm e}^{-2\,\alpha
		\,wt}}{\beta}^{2}-1728\,{{\rm e}^{2\,\alpha\,wt}}\beta-864\,{{\rm e}^{
		2\,\alpha\,wt}}+{{\rm e}^{-4\,\alpha\,wt}}+18\,{\beta}^{4} \right)\times\nb\\
&&{{\rm e}^{-4\,\alpha\,wt}}{ \left[-3(\beta+2)+{{\rm e}^{-2\,
			\alpha\,wt}} \right] ^{-4}},
\eqn
and it is singular at the time
\bq 
\label{ts}
t_{sing}=-\,{\frac {\ln  \left( 3(\beta+2) \right) }{2\alpha\,w}}.
\eq

{It is important to call attention that only in the particular case for $\beta=0$ ($\alpha=\pm 1$) we have a finite value for this scalar, that is,
\bq
\lim_{t \rightarrow t_{sing}} K \rightarrow {\frac {512}{3}}\,{\pi }^{2}{{\it G}}^{2}{\rho_{{0}}}^{2},
\eq
but, in general, for $\beta \ne 0$ ($\alpha=\pm 1$) we get
\bq
\lim_{t \rightarrow t_{sing}} K \rightarrow +\infty.
\eq
}
{Here we have an important difference between the GR and EA theories, which
is not only due to the difference in the coupling constants.
}
{
The differences among our solutions and the classical de Sitter solution ($a(t) \ge 0$ and $t \ge 0$) are clearer seen in
Figure 1, comparing the curves of a(t) for different values of $\beta=\pm 1$ and $\beta=0$ 
(GR limit).}

\subsection{ {Vacuum solutions with $\Lambda=0$} }
The combination of equations (\ref{Gtt}) and (\ref{Tphiphi}), jointly  $\rho_0=0$, gives
{
\bq
a(t)=\sqrt{\frac{-2k}{\beta+2}}t+t_1,
\lb{Bt0}
\eq
where $t_1$ is a constant of integration, $\beta> -2$ and $k \le 0$ {or  $\beta< -2$ and $k \ge 0$} . In the case of $\beta=0$ this represents the Milne cosmological solution \cite{Milne}.
}

For the solution (\ref{Bt0}), using equation (\ref{DEMT}) we obtain
\bq
p=\rho_0 =0.
\eq

The equation (\ref{K}) gives for this solution
{
\bq
K=\frac{3 \beta^2}{(t-t_1)^4}.
\eq
}

Notice that the Kretschmann scalar {does not vanish} unless $\beta=0$. {This result is surprising, without any analogous case in the GR theory, revealing another difference between both theories.} Thus for $\beta\neq 0$ it presents a temporal singularity at
{$t=t_1$} although we do not have any
matter in the spacetime, representing a non-flat vacuum spacetime, {curved by the aether field}. 
{Note that this solution is not a particular case of any previous solutions since if we put $\rho_0=0$ in the equations (\ref{Bt1})-(\ref{Bt3}) the solutions become static.}

{
\section{Analysis of the solutions}
}
In this section we will study the dynamics of the found solutions in two
different modes. There
are two ways to analyze the dynamics: (i) studying only the temporal evolution of the geometrical radius, equation (\ref{Rg}), as done in works of gravitational collapse models (Figure \ref{fig2}); (ii) or using directly the scale factor $a(t)$ as done in cosmological model works (Figures \ref{fig1}, \ref{fig3},
\ref{fig4} and \ref{fig5}).

Thus, the dynamic of the solutions, obtained in Section 3, will be realized by the study of the time evolution of scale factor and its derivatives, represented by the Hubble expansion rate
\begin{equation}
\label{Ht}
H(t)=\frac{\dot{a}(t)}{a(t)},
\end{equation}
the deceleration parameter, defined by
\bq
q(t) = -\frac{a(t)\ddot{a}(t)}{\dot a^2(t)}
\lb{qt}
\eq
{and the geometrical radius defined by}
\bq
R_{ge}=r|a(t)|.
\lb{Rg}
\eq

The parameters $a(t)$, $H(t)$ and $q(t)$ are useful in the comparison of EA solutions with the FLRW solutions of GR, allowing us to make explicit the differences and similarities between the two theories.

\subsection{{Dynamics of the vacuum solutions with $\Lambda>0$} }

\begin{figure}[!ht]
\centering
\includegraphics[width=7cm]{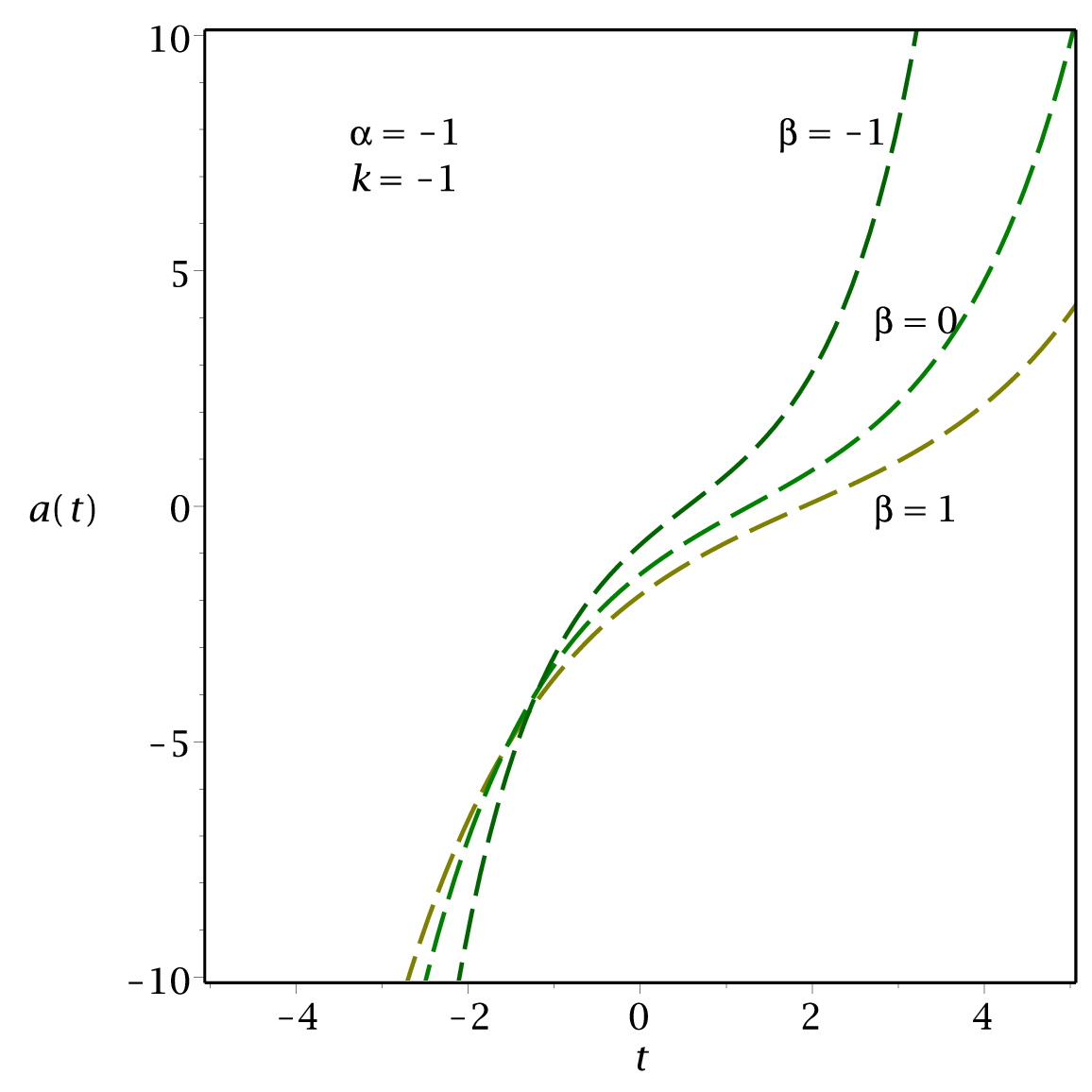}
\includegraphics[width=7cm]{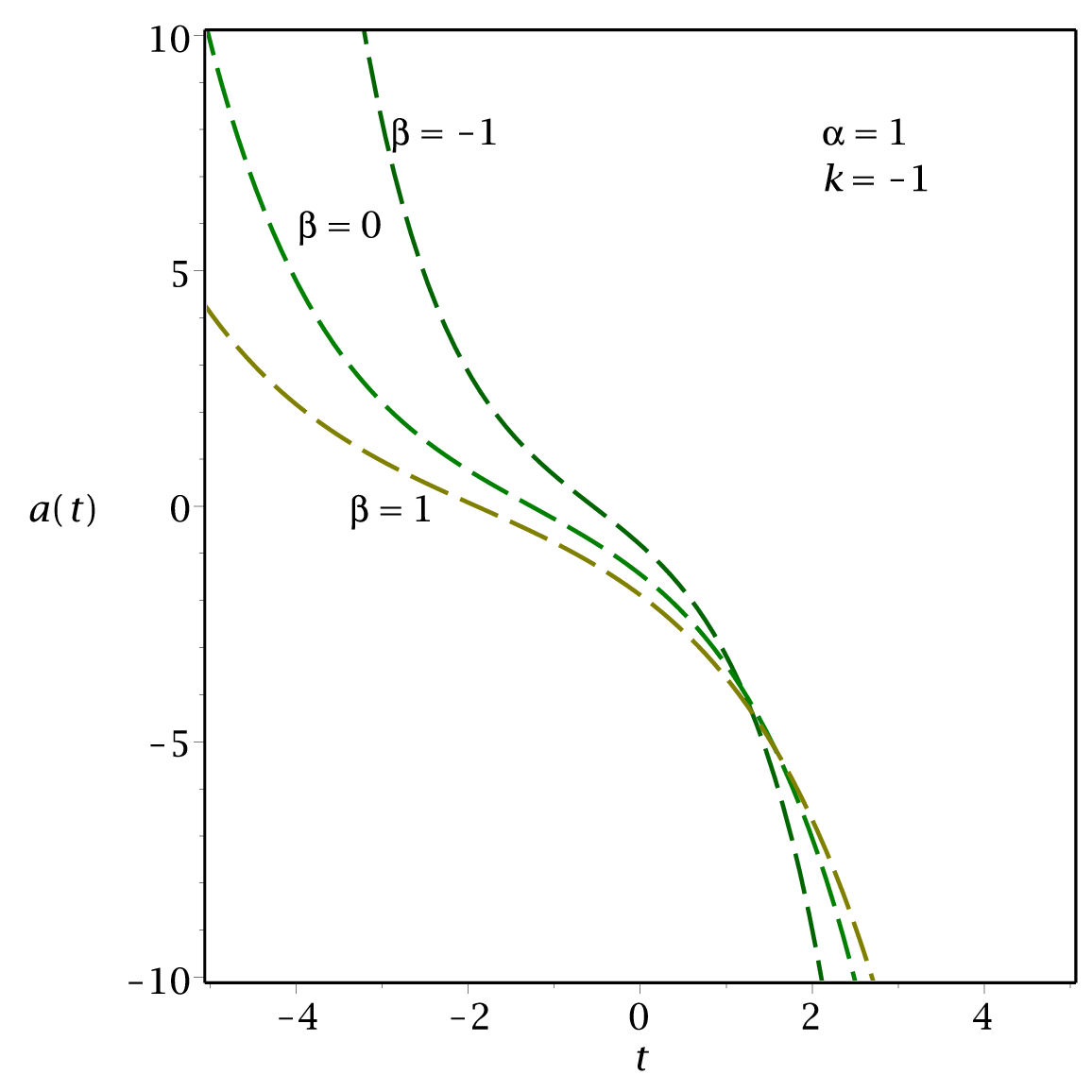}
\includegraphics[width=7cm]{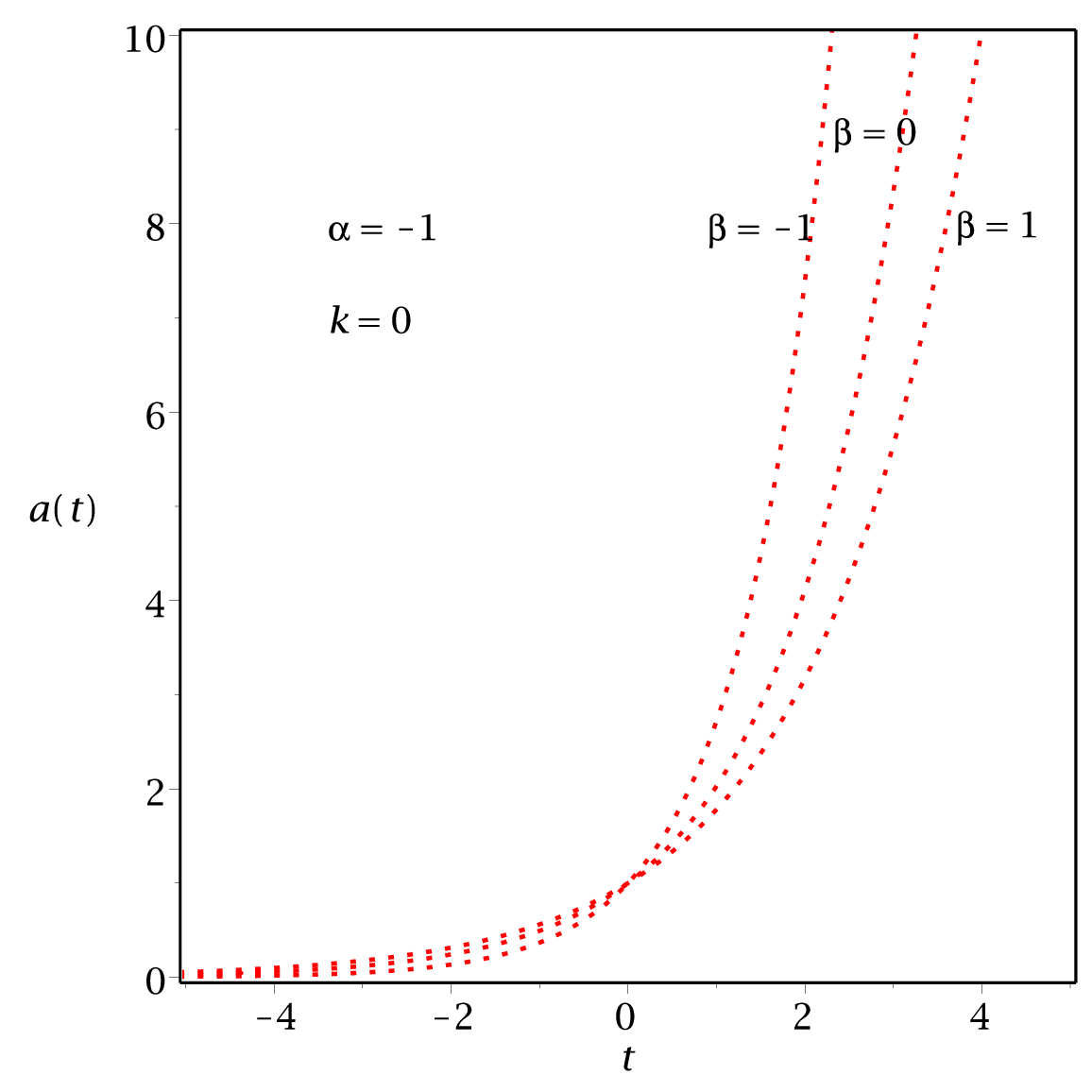}
\includegraphics[width=7cm]{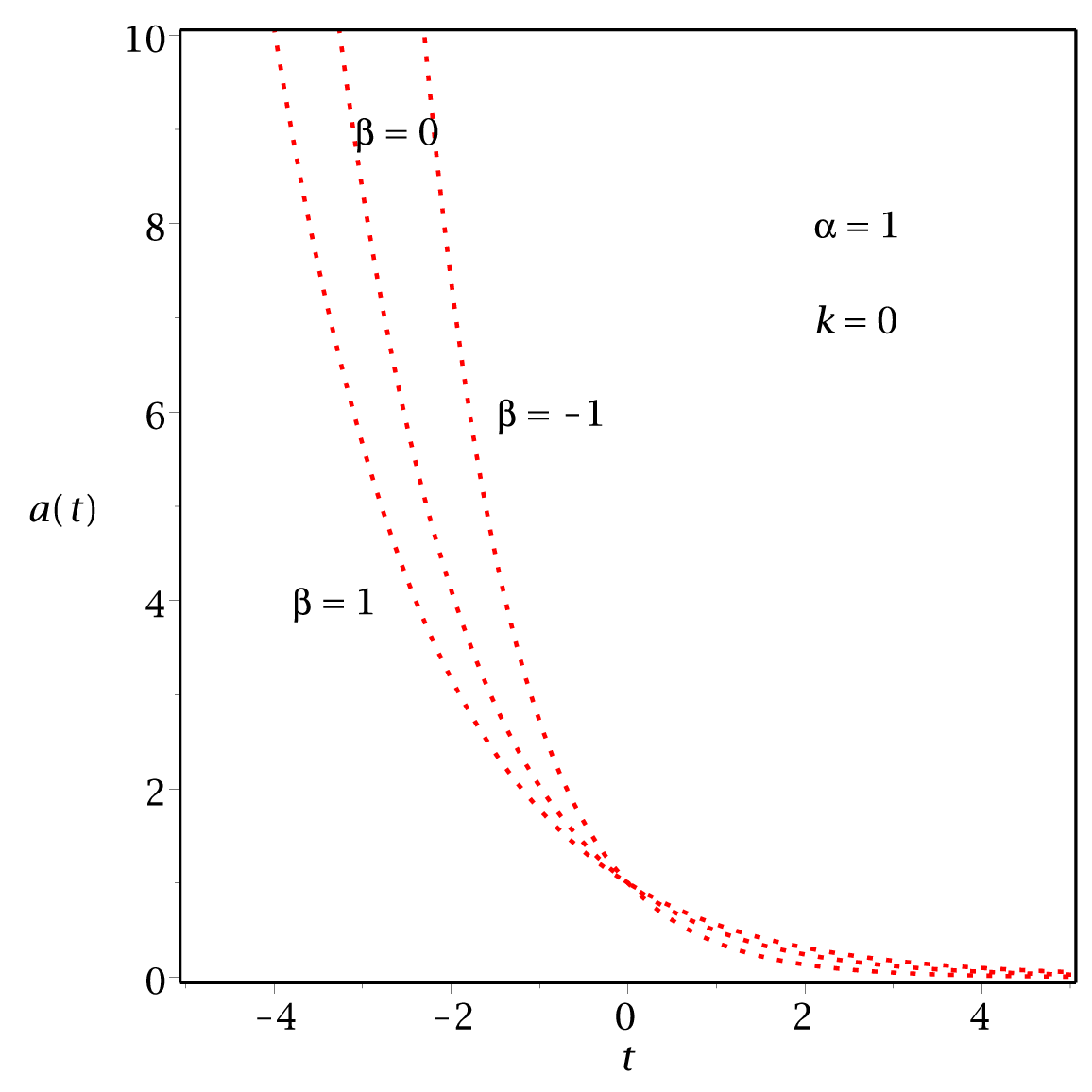}
\includegraphics[width=7cm]{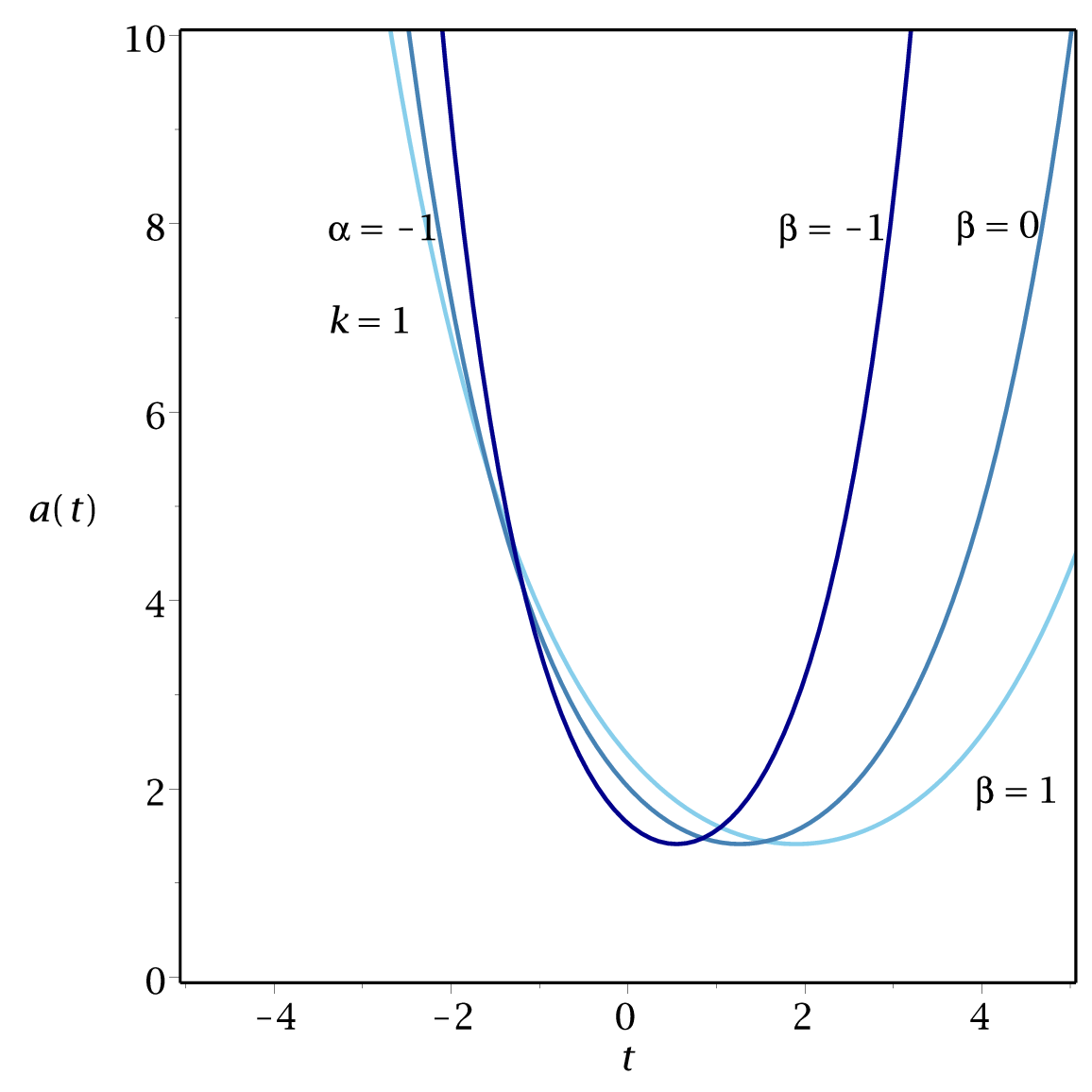}
\includegraphics[width=7cm]{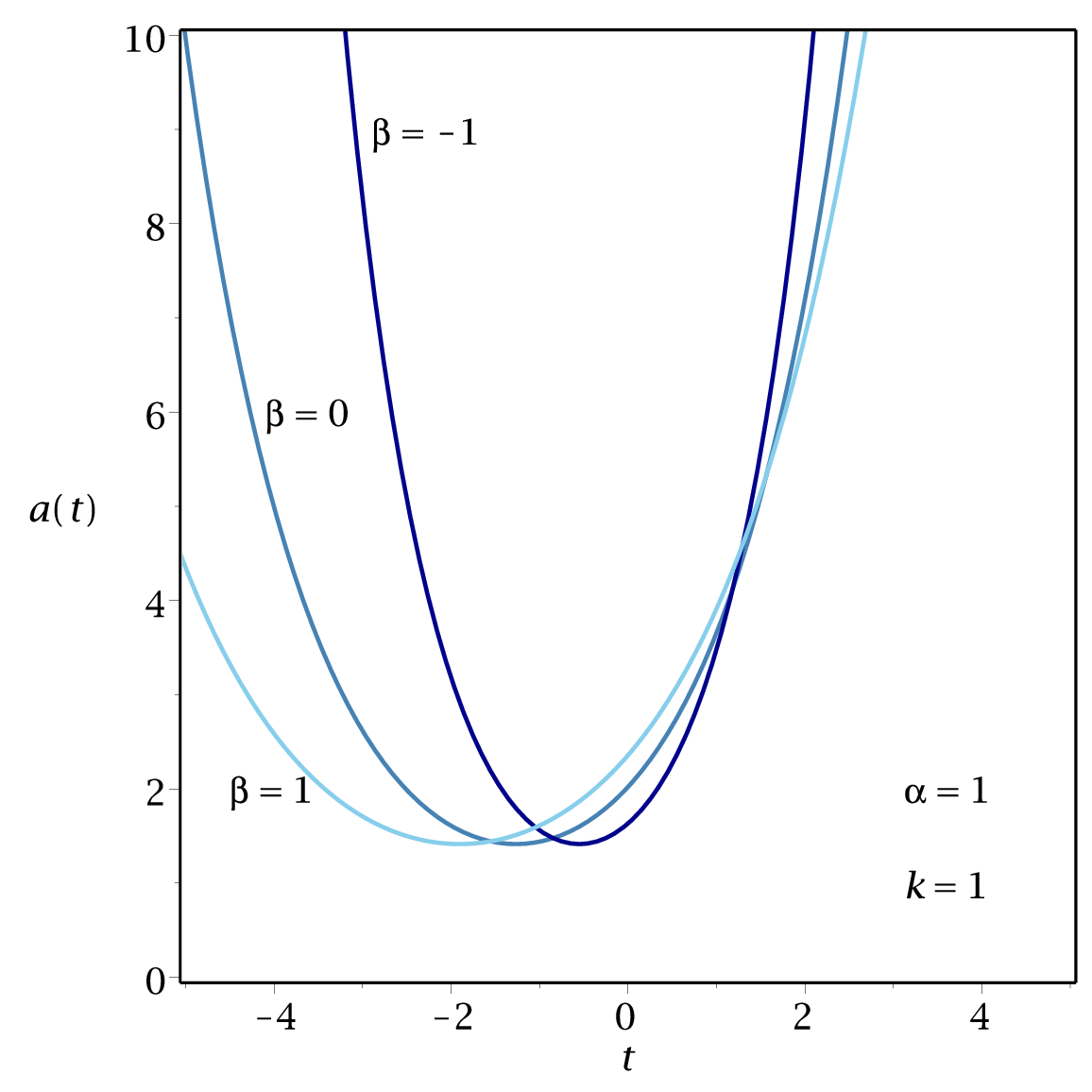}
\caption{Scale factor evolution in time, derived for a fluid with constant energy density, equations (\ref{Bt1})-(\ref{Bt3}), for different values of $k=-1, 0, 1$, $\beta=-1,0,1$, and $\alpha= \pm 1.$}
\label{fig1}
\end{figure}

The scale factor {for the solutions} (\ref{Bt1})-(\ref{Bt3}), derived for the case in which $\rho=\rho_0$, are shown in the Figure \ref{fig1}, for $k=-1, 0, 1$, {$\beta=-1,0,1$} and $\alpha=\pm 1$. For the sake of simplicity, it was adopted that the constant factor in the parameter $w$, given in (\ref{omega}), assumes an unit value, i.e., $4\sqrt{\frac{\pi G \rho_0}{3}}=1$, in all the figures.  

\begin{figure}
\centering
\includegraphics[width=7cm]{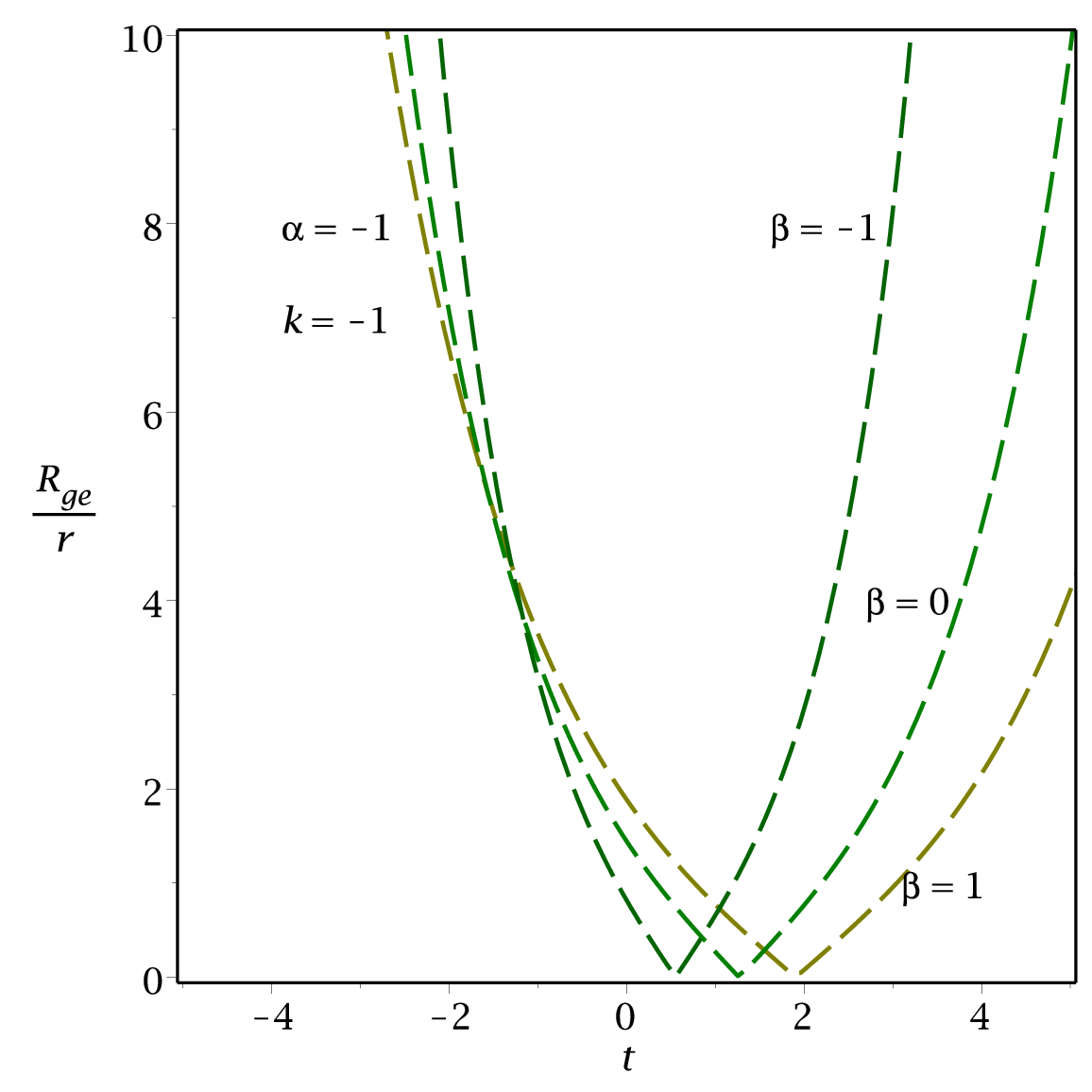}
\includegraphics[width=7cm]{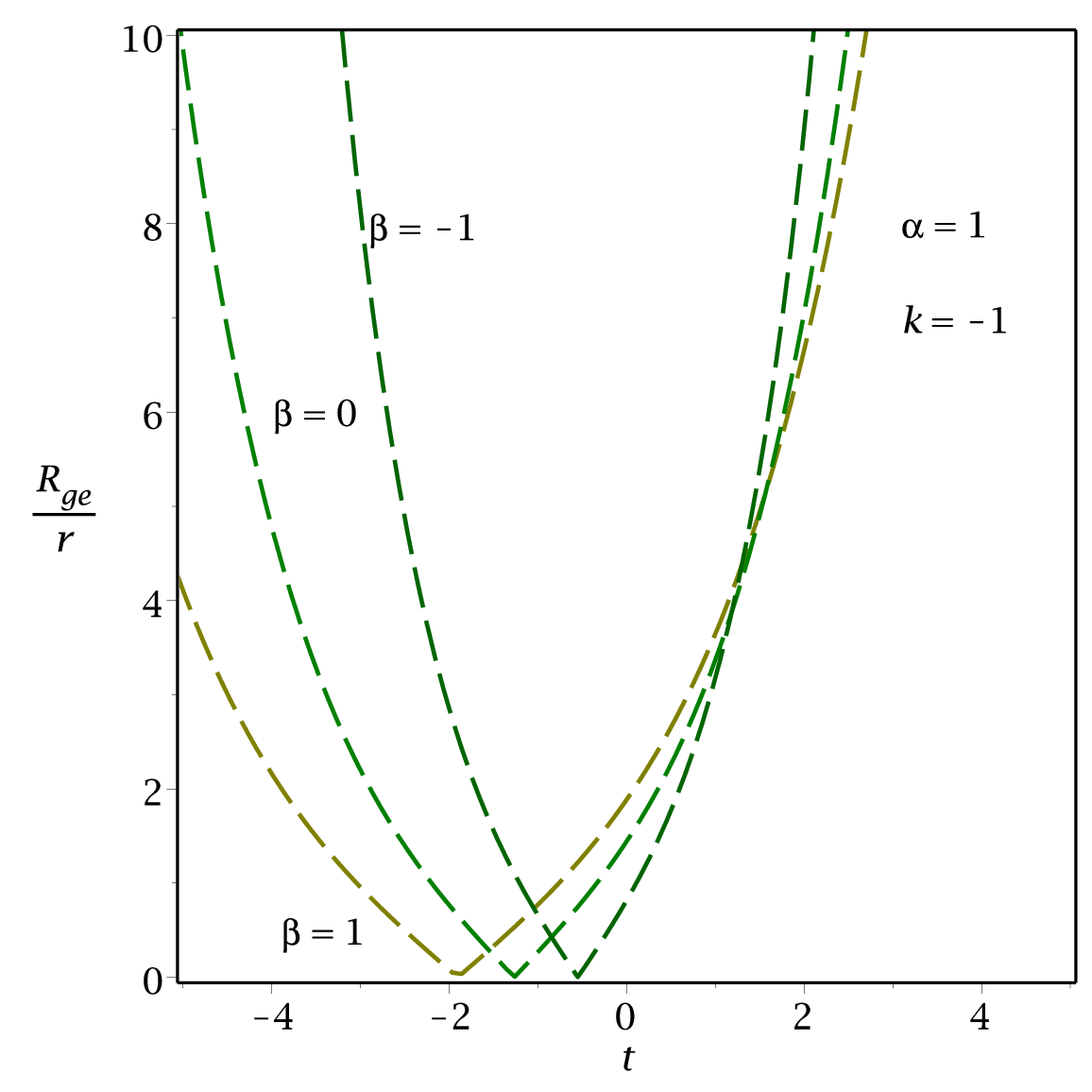}
\includegraphics[width=7cm]{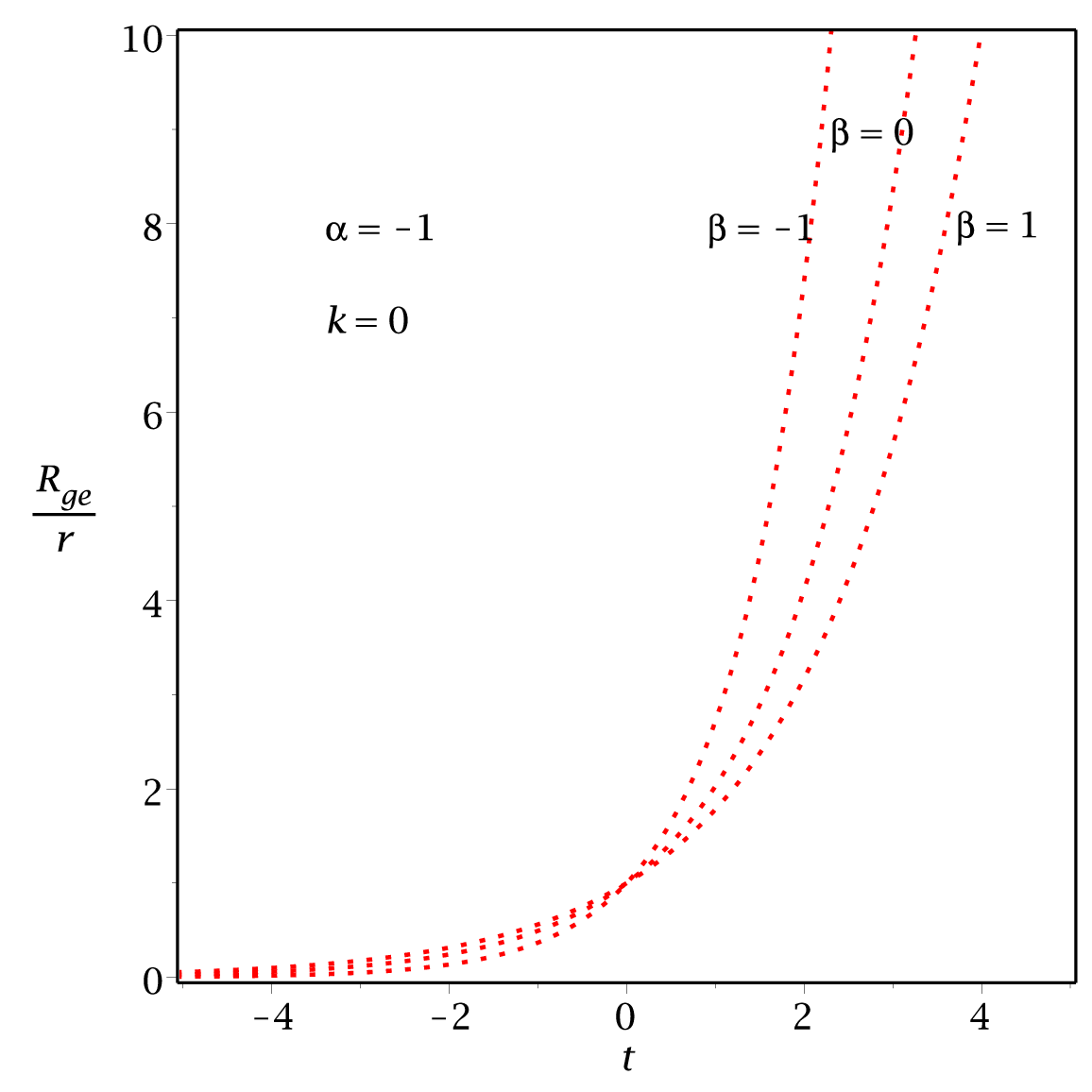}
\includegraphics[width=7cm]{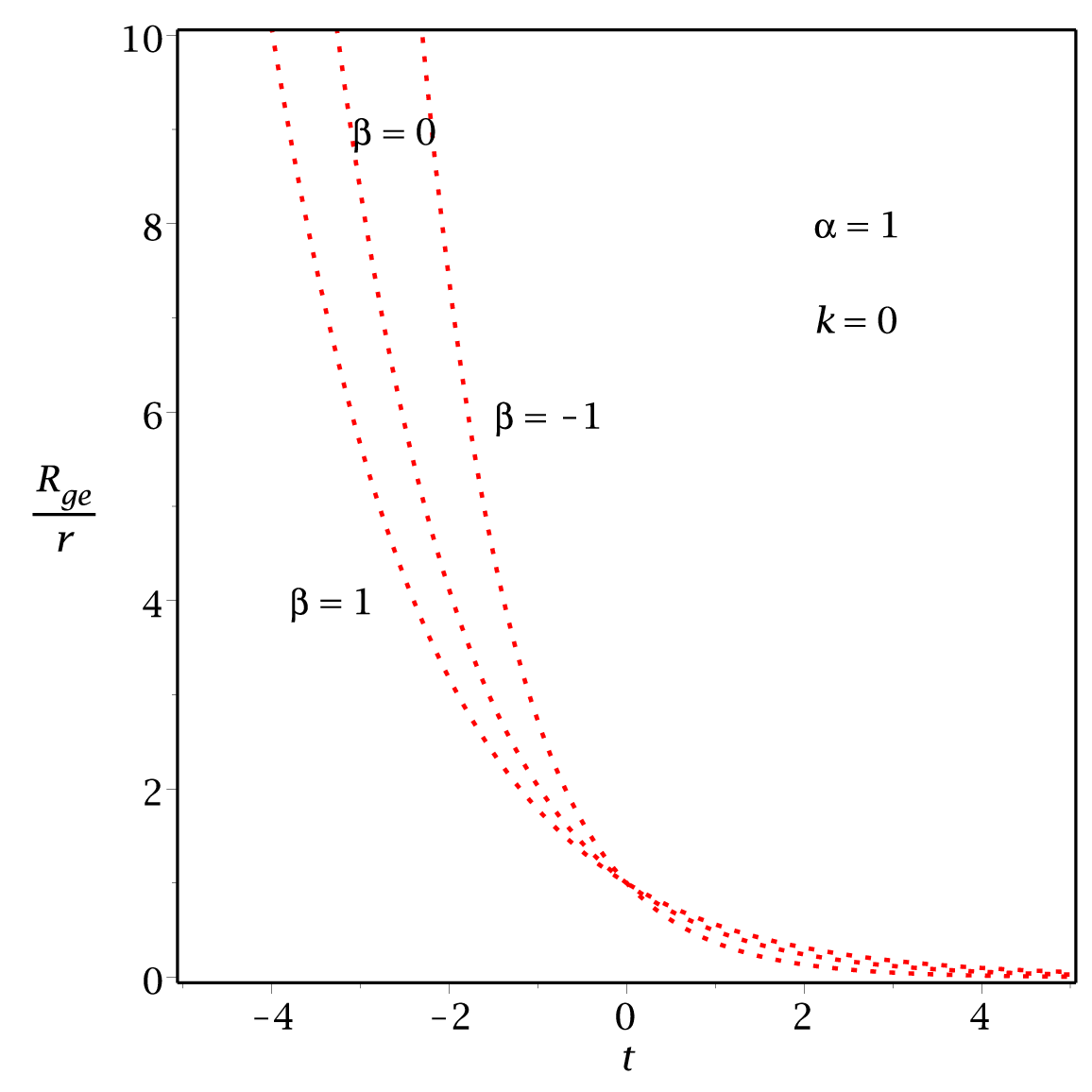}
\includegraphics[width=7cm]{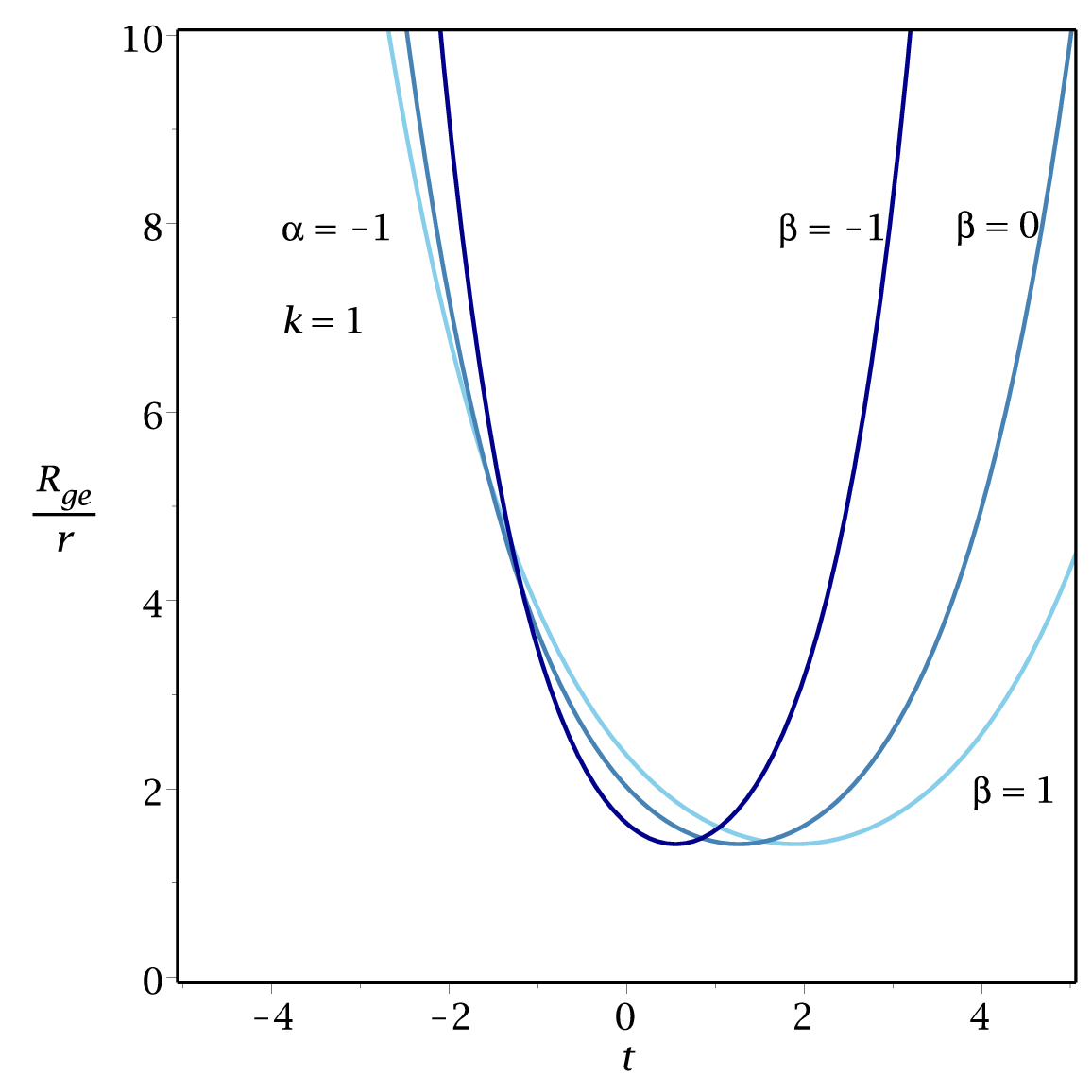}
\includegraphics[width=7cm]{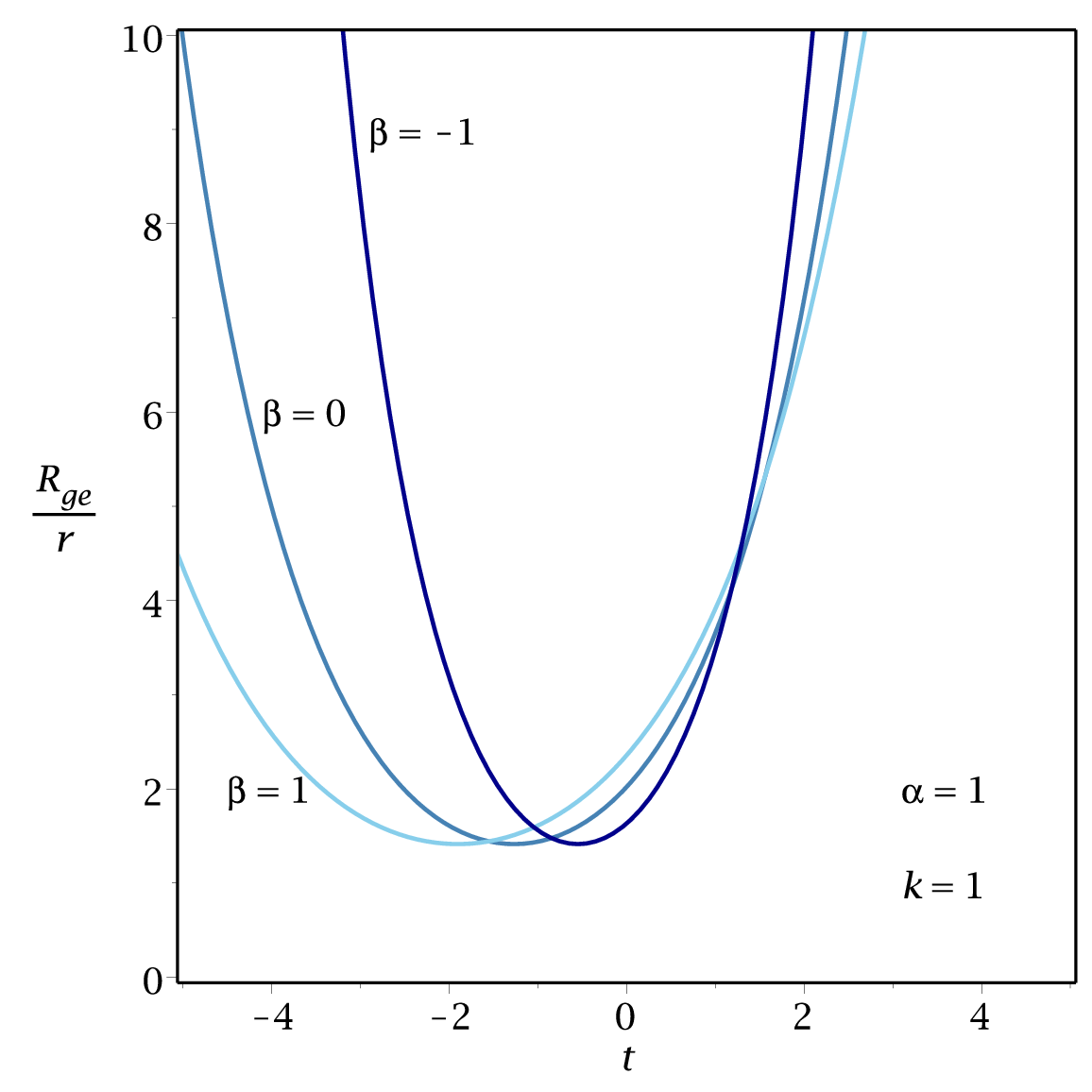}
\caption{Geometric radius evolution in time, derived for a fluid with constant energy density, equations (\ref{Bt1})-(\ref{Bt3}), for different values of $k=-1, 0, 1$, $\beta=-1,0,1$, and $\alpha=\pm 1.$}
\label{fig2}
\end{figure}

\begin{figure}
\centering
\includegraphics[width=7cm]{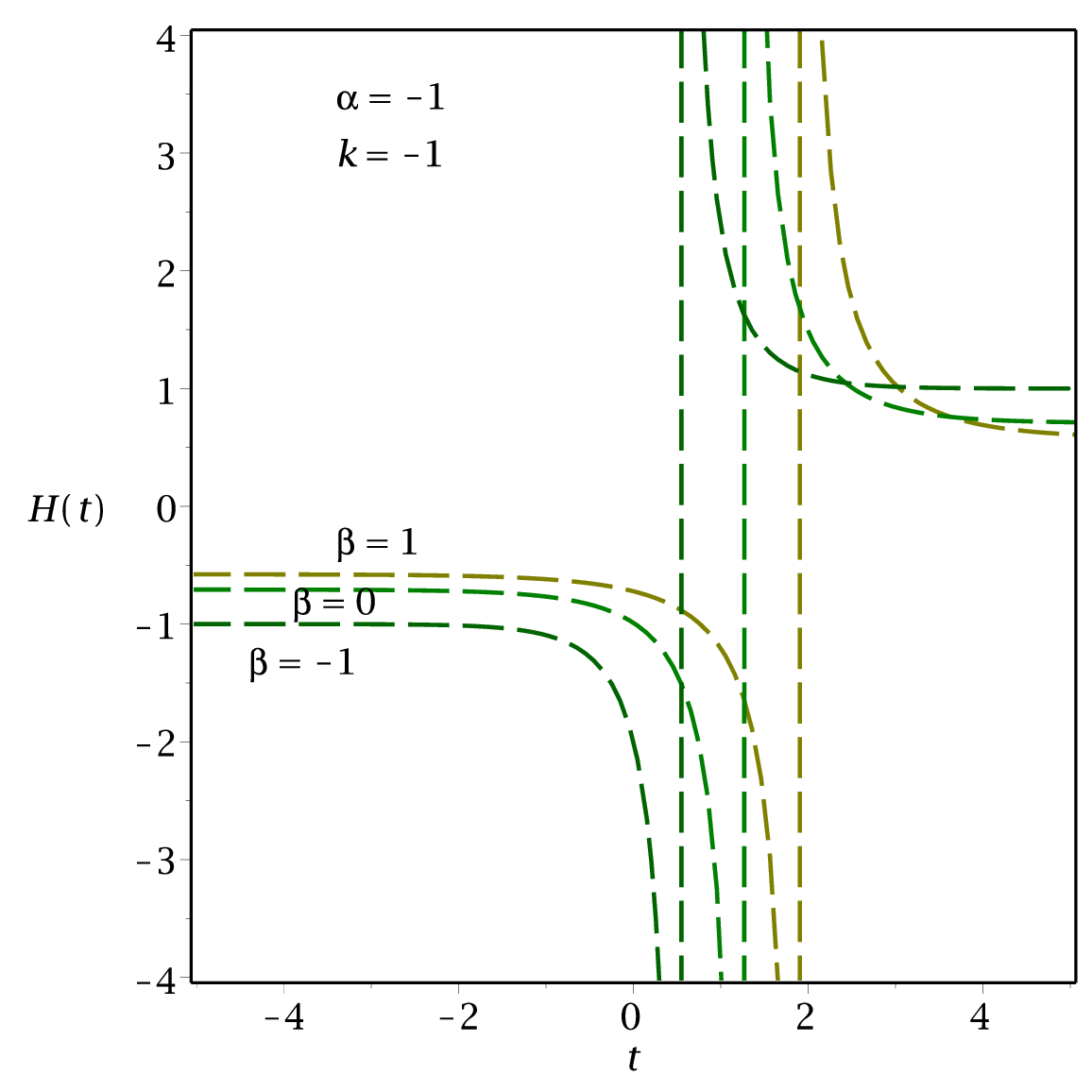}
\includegraphics[width=7cm]{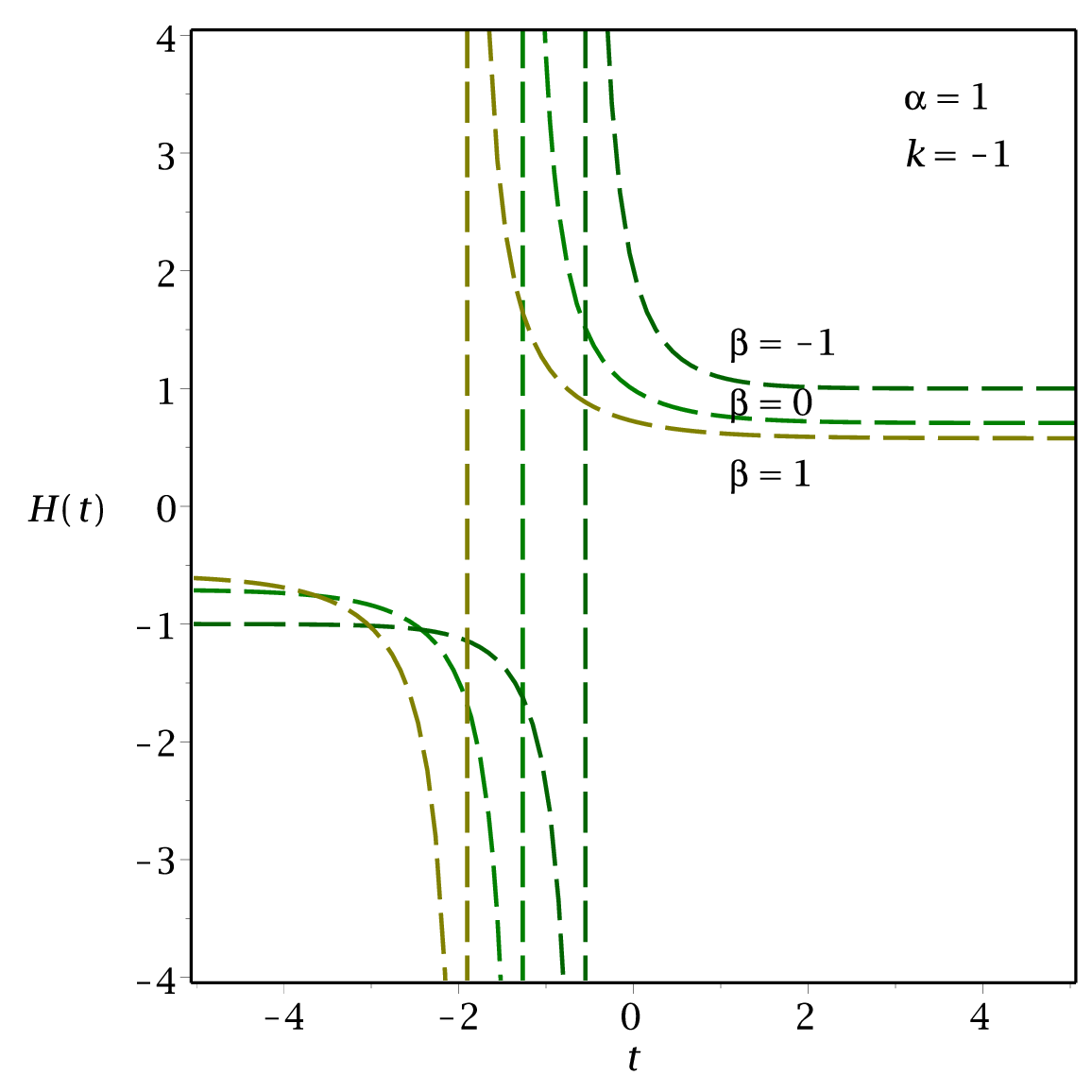}
\includegraphics[width=7cm]{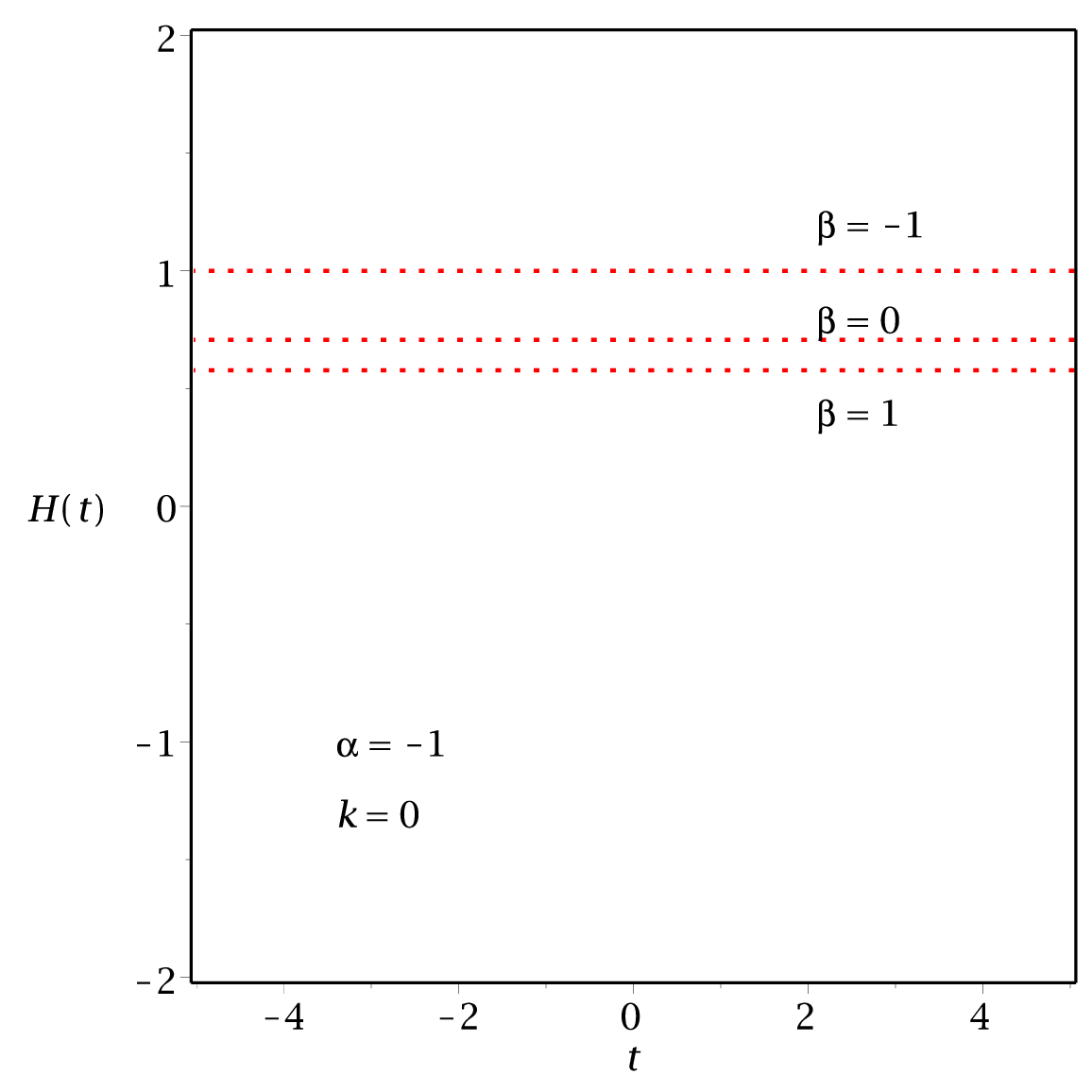}
\includegraphics[width=7cm]{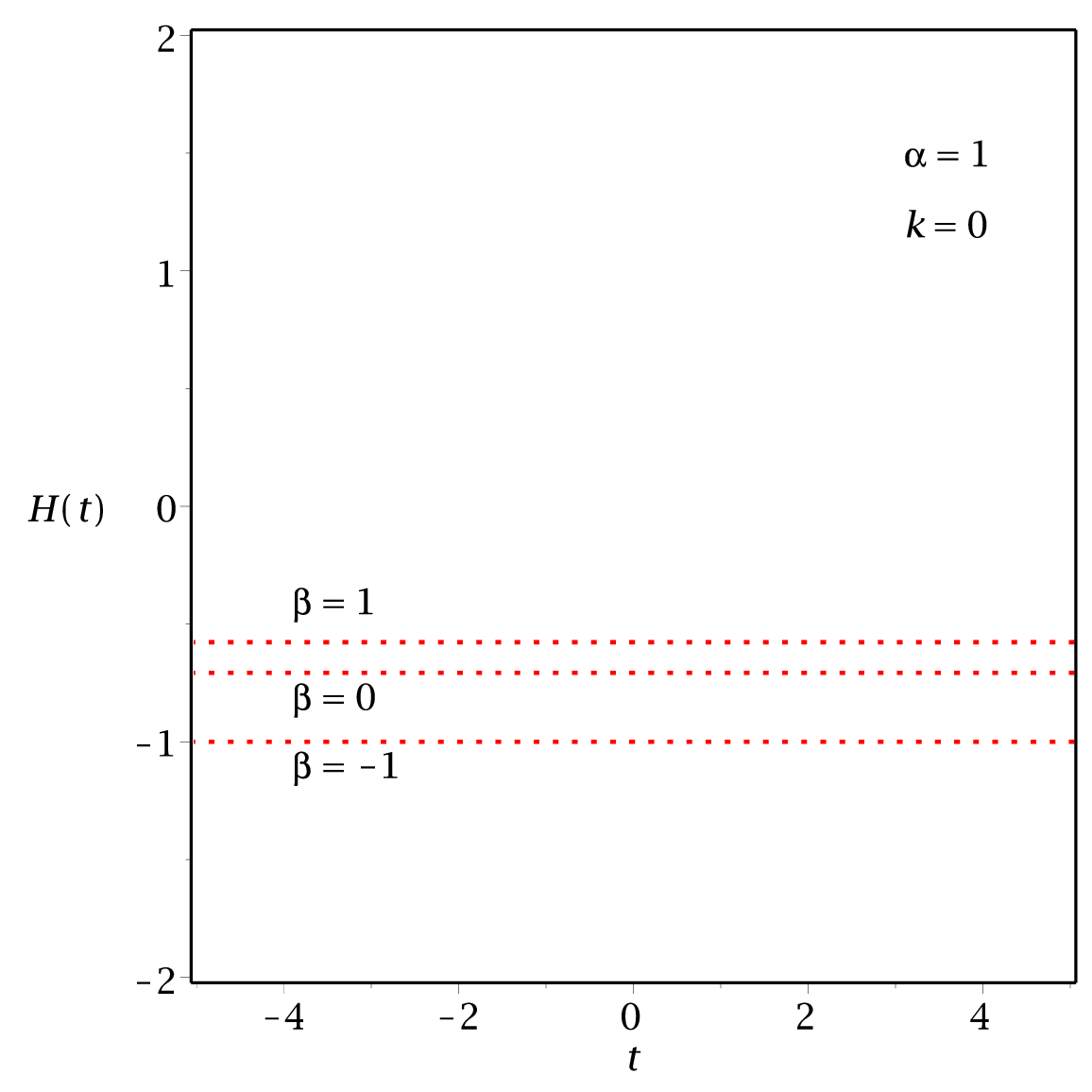}
\includegraphics[width=7cm]{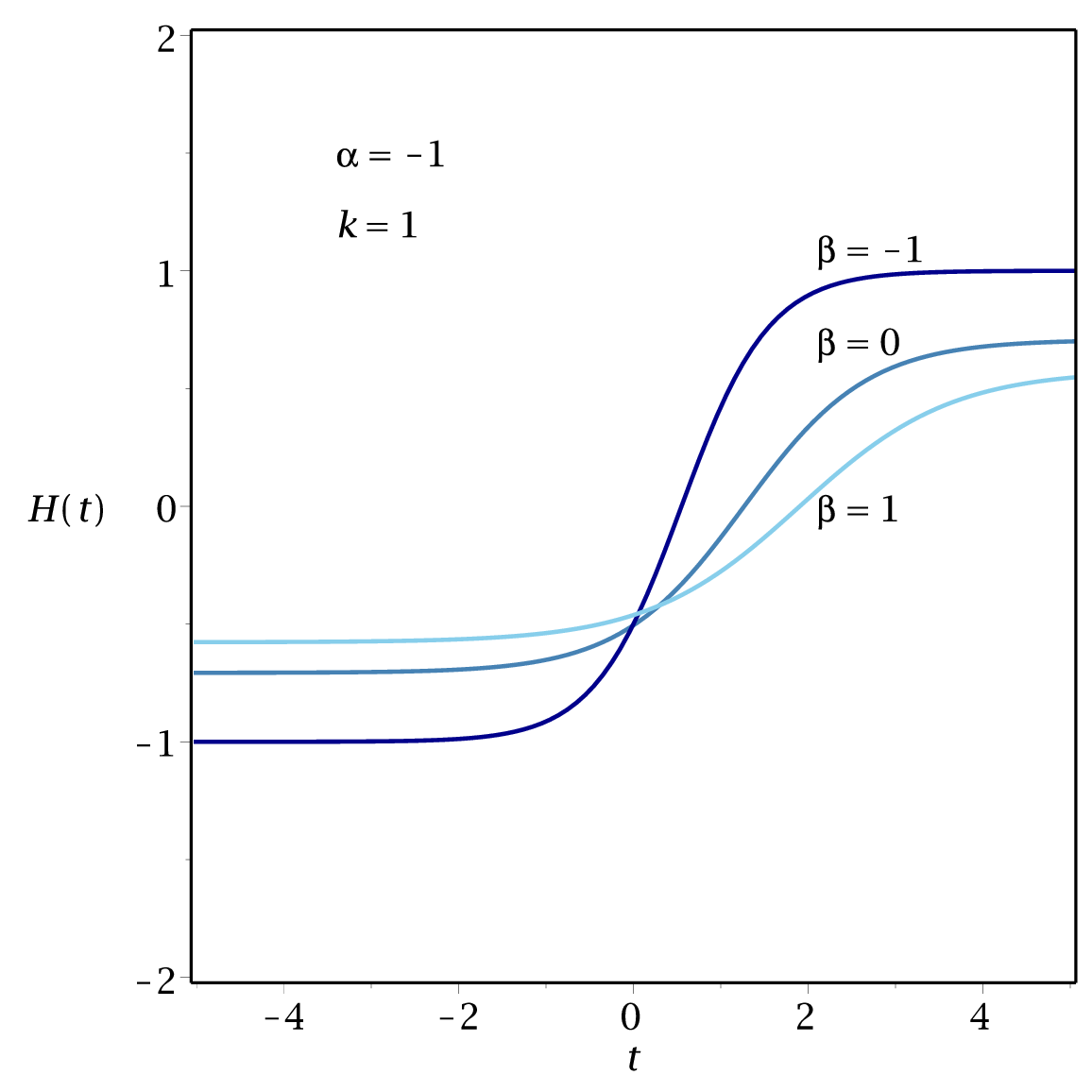}
\includegraphics[width=7cm]{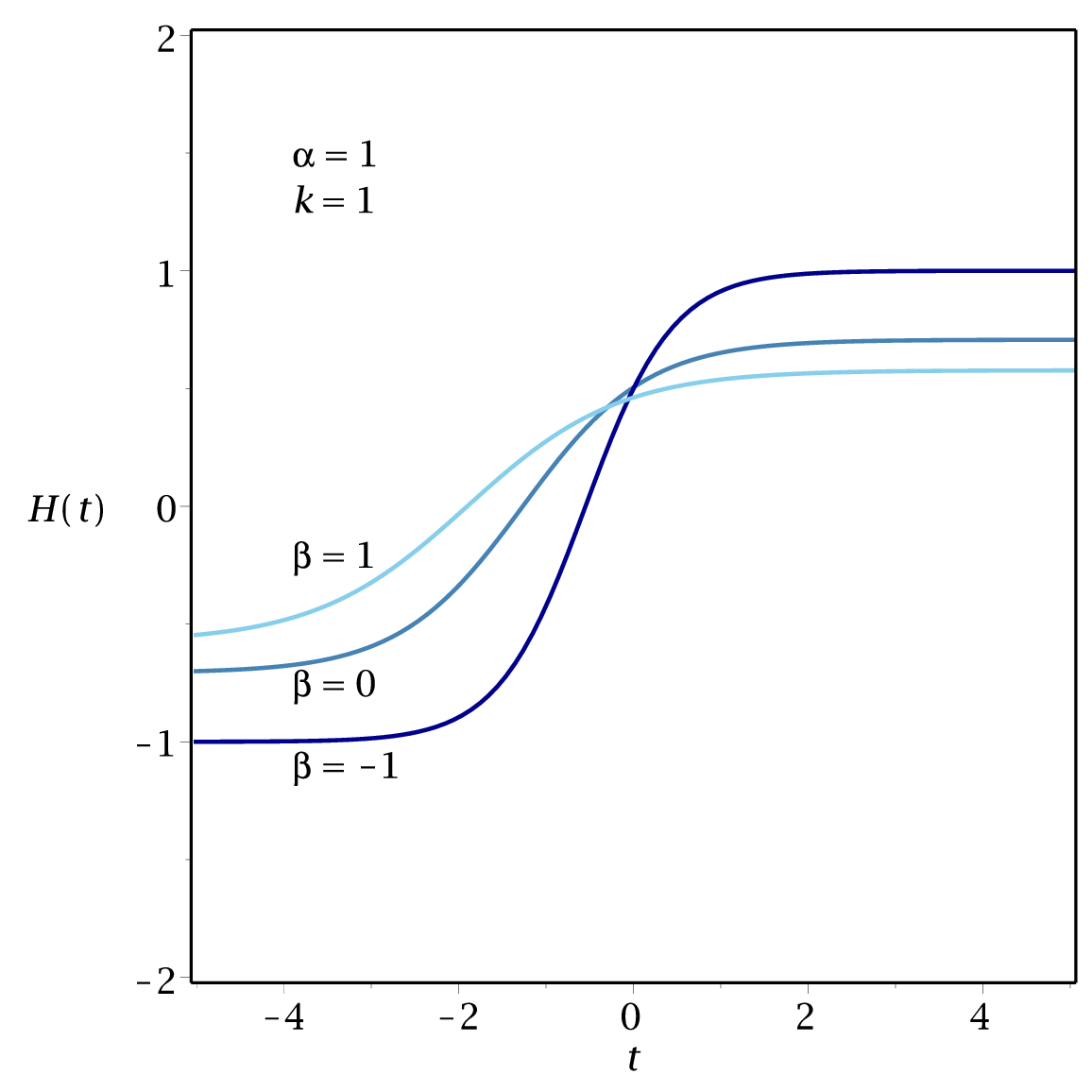}
\caption{Hubble expansion rate, derived for a fluid with constant energy density, equations (\ref{Bt1})-(\ref{Bt3}), for different values of $k=-1, 0, 1$, $\beta=-1,0,1$, and $\alpha= \pm 1.$}
\label{fig3}
\end{figure}

\begin{figure}
\centering
\includegraphics[width=7cm]{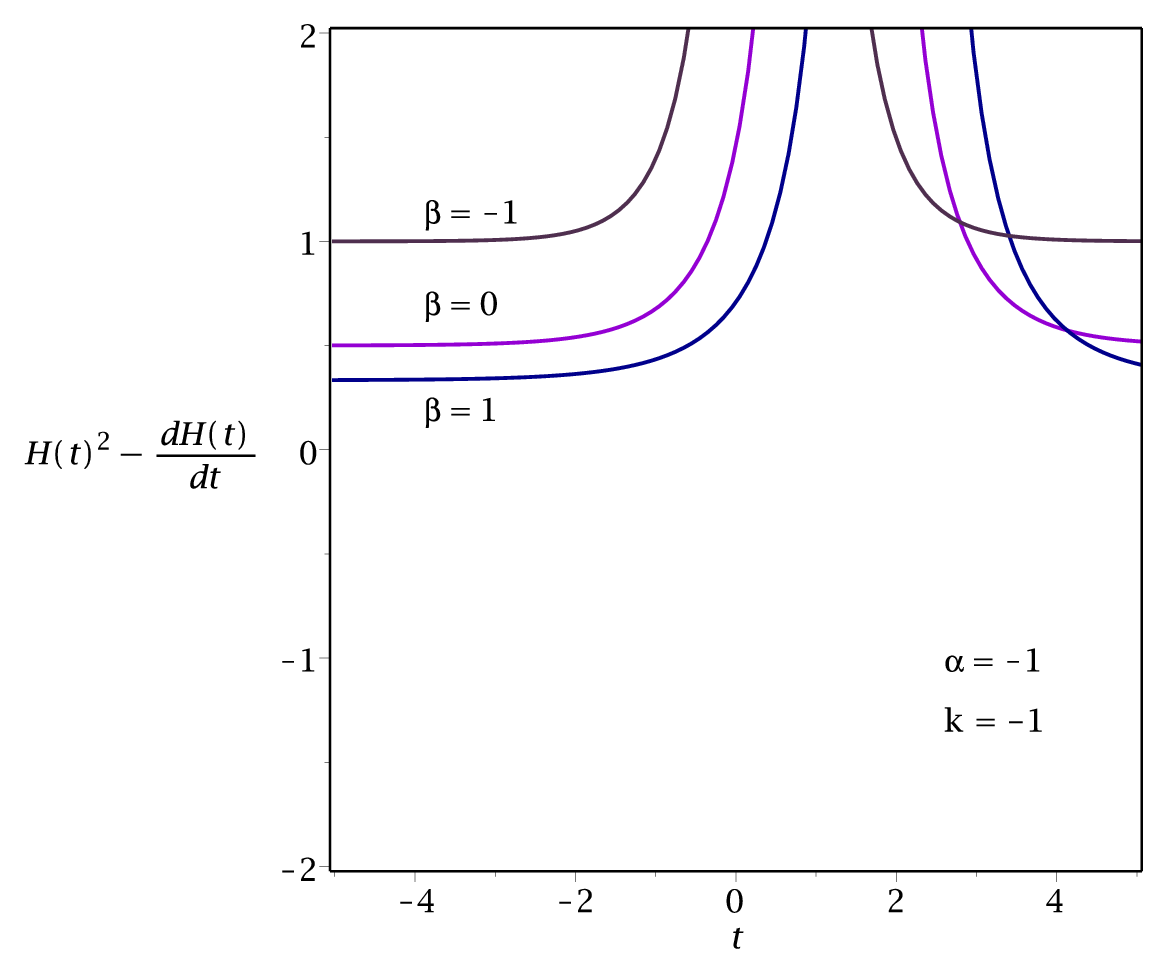}
\includegraphics[width=7cm]{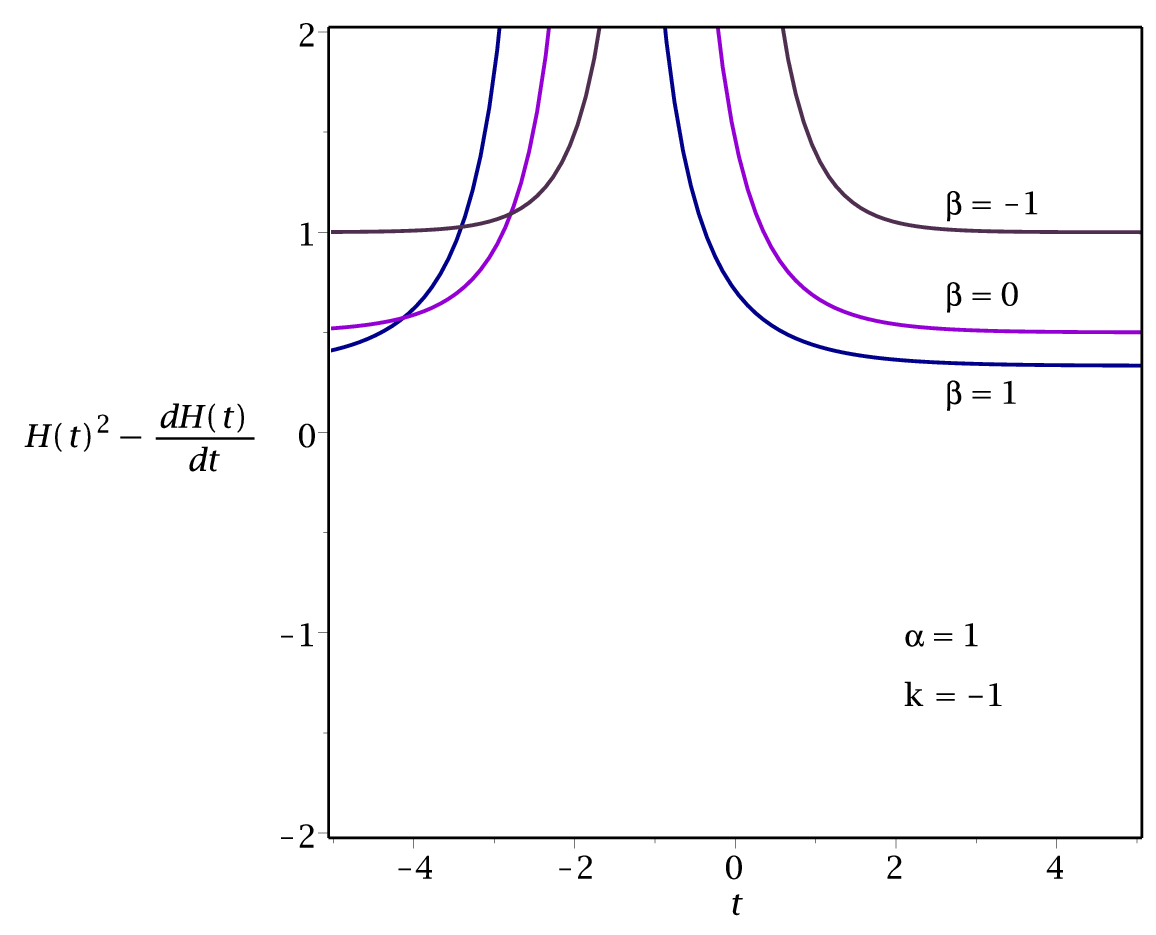}
\includegraphics[width=7cm]{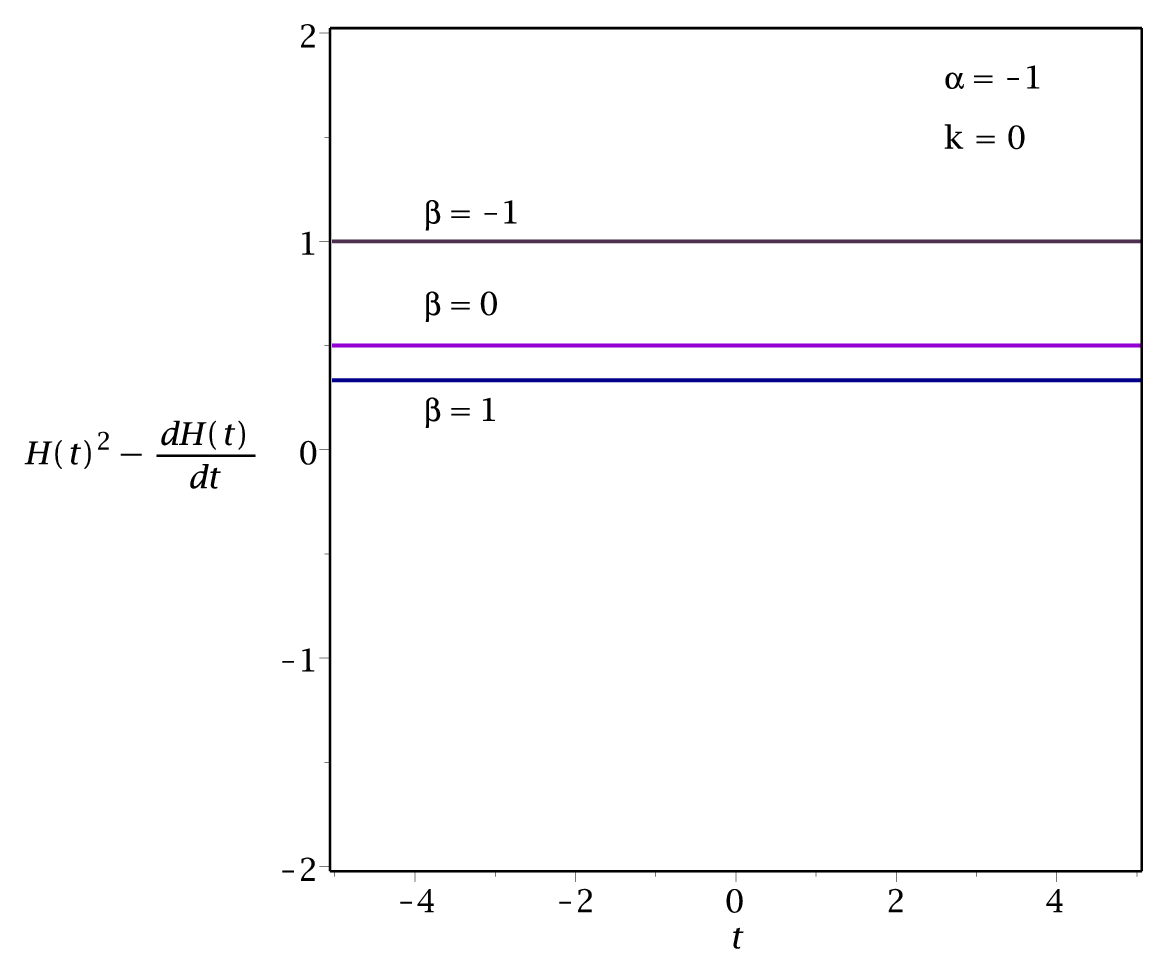}
\includegraphics[width=7cm]{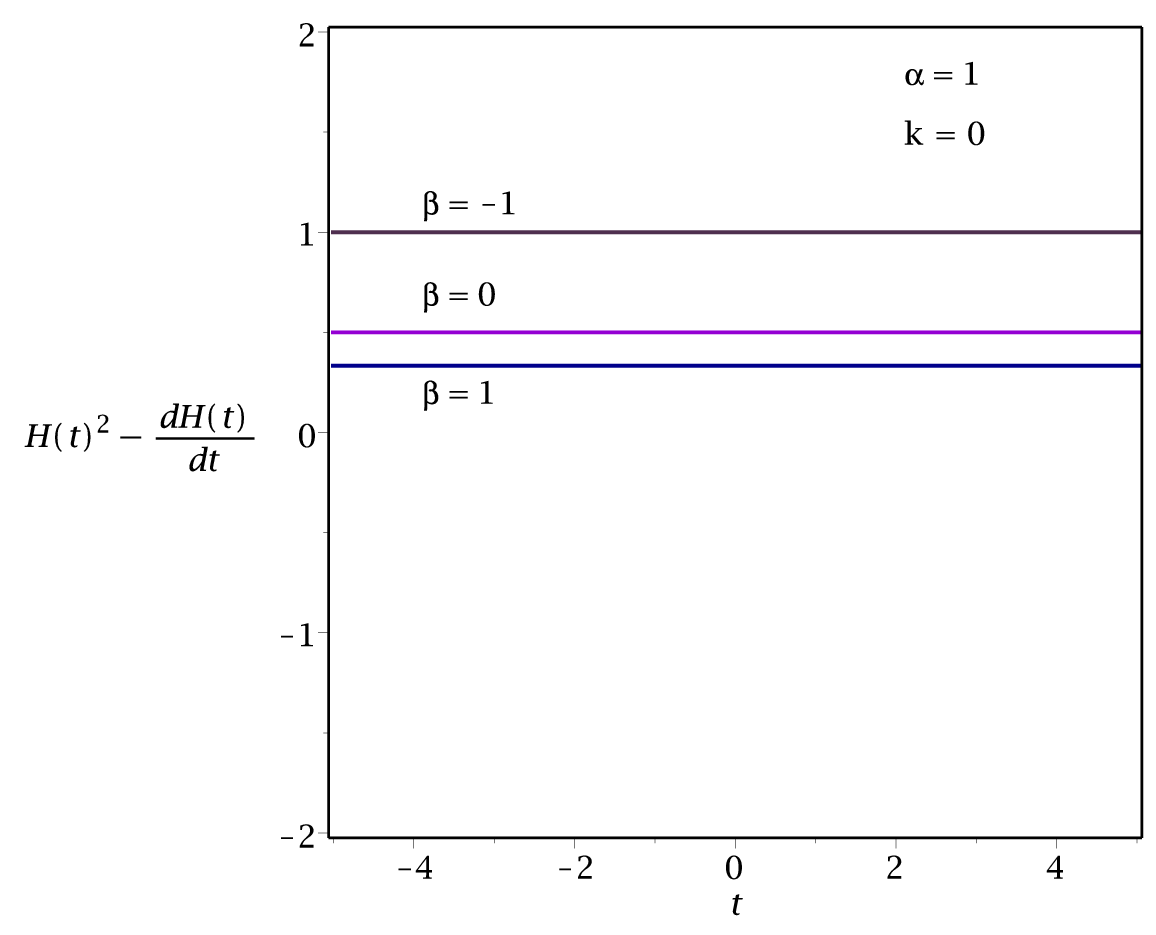}
\includegraphics[width=7cm]{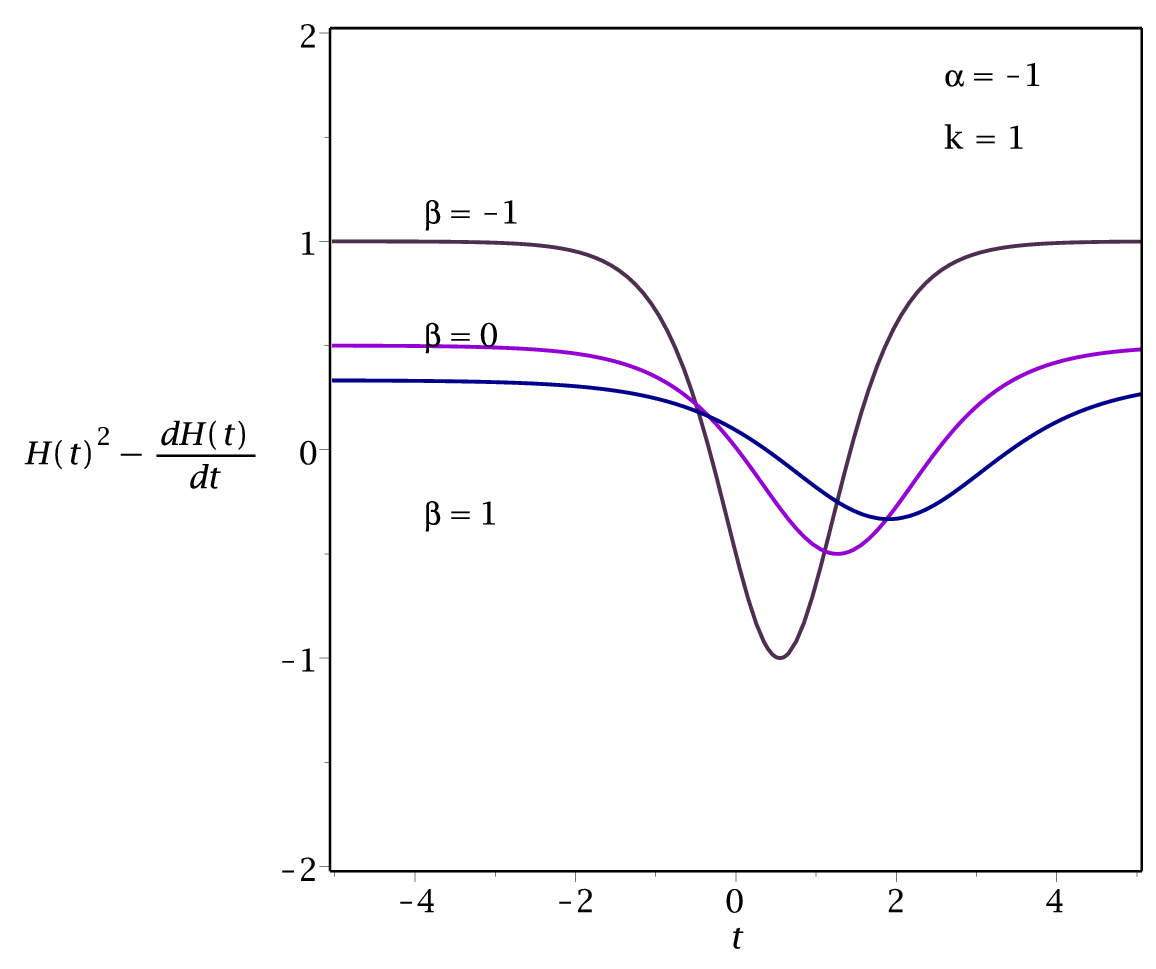}
\includegraphics[width=7cm]{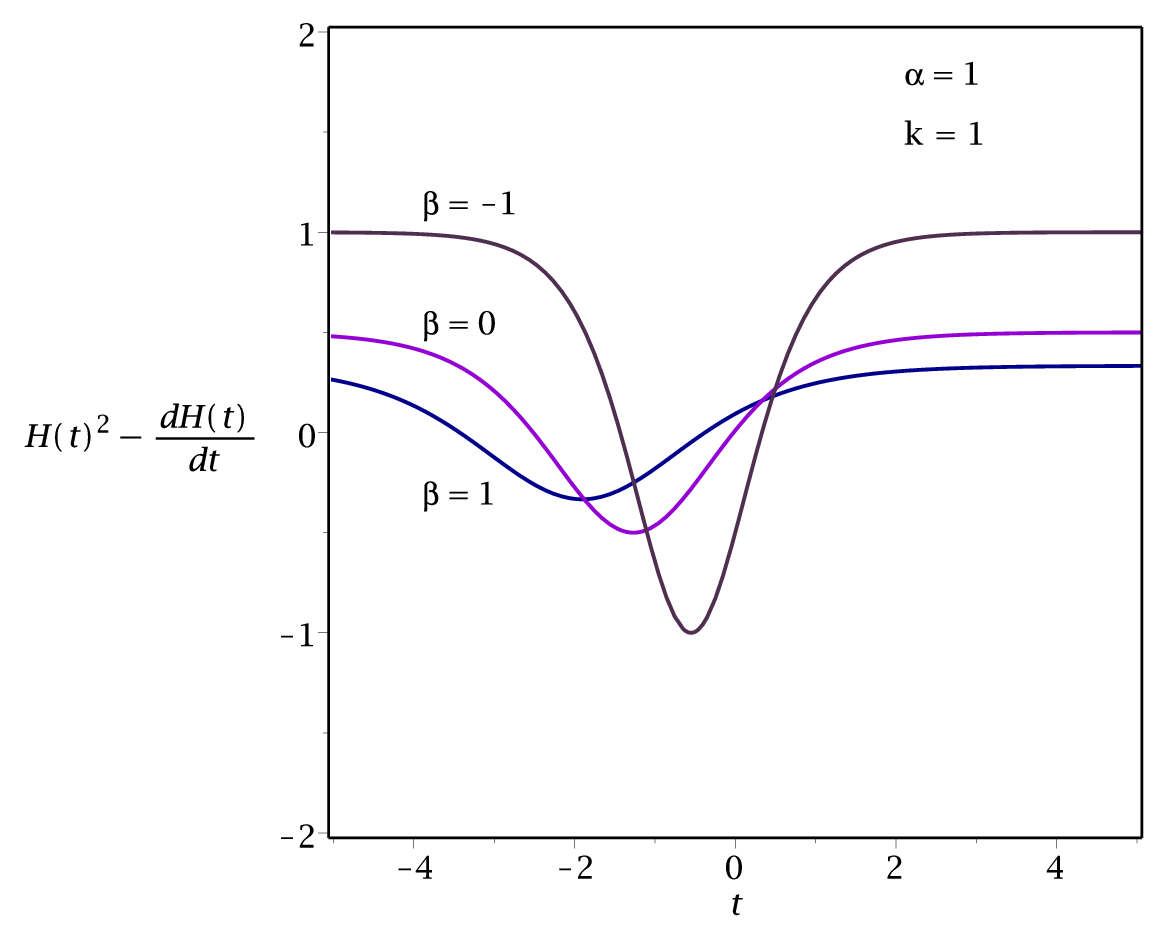}
\caption{The quantity $(H^2-\dot{H})$, for a fluid with constant energy density, for  for different values of $k=-1, 0, 1$, $\beta=-1,0,1$, and $\alpha= \pm 1.$}
\label{fig4}
\end{figure}

Alternatively, it can be represented in terms of geometric radius $R_{ge}=r|a(t)|$ as show the Figure \ref{fig2}, where the negative part of scale factor is suppressed by the calculation of its absolute value.
In this figure we can see clearly that the the system with $k=-1$ 
{($\beta \ne 0$ and $\alpha=\pm 1$)} collapses
to a singularity at the time $t_{sing}$, given by the equation 
(\ref{ts}), and after expands. For $k=0$ the system always expands ($\alpha=-1$) or contracts ($\alpha=1$). However, for $k=1$ the system
contracts to a minimum radii and after expands, clearly showing a bouncing 
effect as in some gravitational collapse models \cite{Herrera}.
If we try to analyze the same bouncing effect in a cosmological model
using the scale factor instead the geometrical radius, we must calculate
the Hubble expansion rate $H$ and the quantity $H^2-dH/dt$ (see Figures
\ref{fig3} and \ref{fig4}).

The Hubble expansion rate, defined in (\ref{Ht}), is shown for each one of the solutions in Figure \ref{fig3}. {Note that {for $k \neq 0$}, the effect of the vector field is not simply to change the value of the coupling constant. {It is important to solve} the field equations in order to make explicit the new function $a(t)$ which defines the scaling factor and, consequently, the expansion factor $H(t)$. In fact, these functions differ from those in the GR theory, being modified by the vector field. As shown in Figure 3, the expansion rate may decrease as predicted by Carroll and Lim, but may also increase, when compared to that expected in the GR theory ($\beta=0$), depending on the choices of the parameters $\beta$ and $\alpha$. In particular, we can see that this behavior is reversed when we exchange the $\beta$ sign, or $\alpha$ sign.}

{Through the figures, it is possible to notice that the $\beta$ variation introduces a small variation in the scale factor and consequently in the other associated parameters. For this reason, in our analysis, we are taking account only the variation of $k$ and $\alpha$, that produce more expressive variations in the results.}

\begin{itemize}
\item {k=-1}\\ 
This case presents a singular point in $t_{sing}=-\ln(6+3\beta)\sqrt{\beta+2}/(2\alpha)$, as shown in Figure \ref{fig2} and it corresponds to the inflection point exhibited in Figure (\ref{fig1}). The value of singular point increases with the $\beta$ value.

{The Figure \ref{fig3}, shows a divergence in the expansion rate in $t_{sing}=- \ln(6+3\beta)\sqrt{\beta+2}/(2\alpha)$. Its divergence coincides with (\ref{ts}), where the Kretschmann is singular.}
This divergence is due to the fact that the expansion factor becomes negative at a certain time. {In most of cosmological models this does not
occur because it is always assume that $a(t) > 0$ for $t > 0$. The signal of $\alpha$, parameter present in the exponential factor in the solutions, only acts to mirror the curves around $t=0$.}

\item{k=0}\\
According to (\ref{Bt2}), the scale factor increases or decreases exponentially, for $\alpha=-1$ or to $\alpha=+1$, respectively. This result is particularly interesting, since it resembles a de Sitter solution in GR, where, for each value of $\beta$, the constant parameter $w$ plays the role of a cosmological constant.

\item{k=1}\\
The existence of a bounce is confirmed by the change of signal in $H(t)$, Figure \ref{fig3}. { Another important characteristic is the signal of the quantity $(H^2-dH/dt)$, what is given by $\rho+p$. During the bounce, it values becomes negative as show Figure \ref{fig4}. }

\end{itemize}

In respect to the acceleration, since (\ref{qt}) is defined to be negative, reason by what is named deceleration parameter. Then, $q(t)< 0$ correspond to an accelerated expansion while $q(t)>0$, corresponds to the opposite case, an decelerated expansion. 

\begin{figure}[!ht]
\centering
\includegraphics[width=7cm]{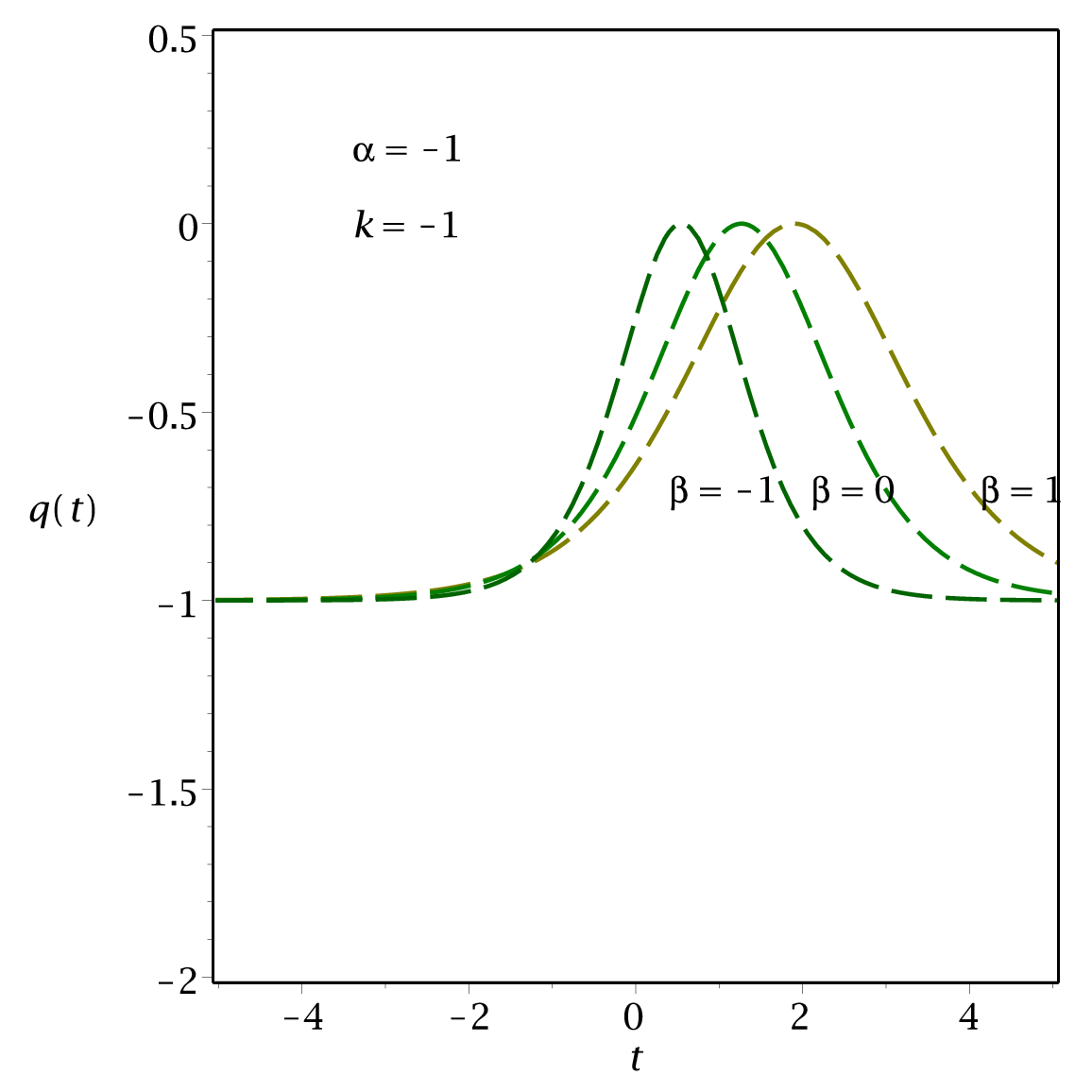}
\includegraphics[width=7cm]{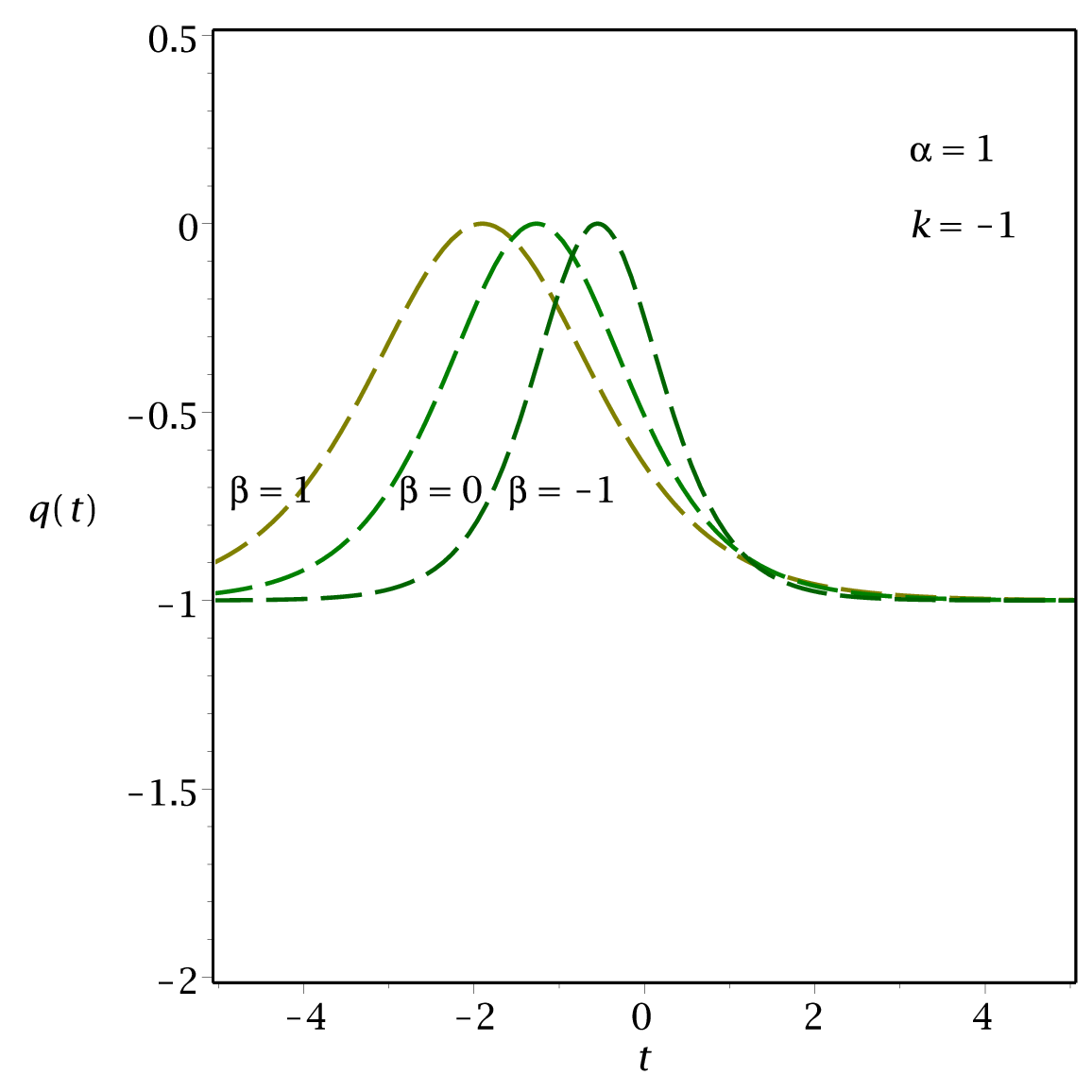}
\includegraphics[width=7cm]{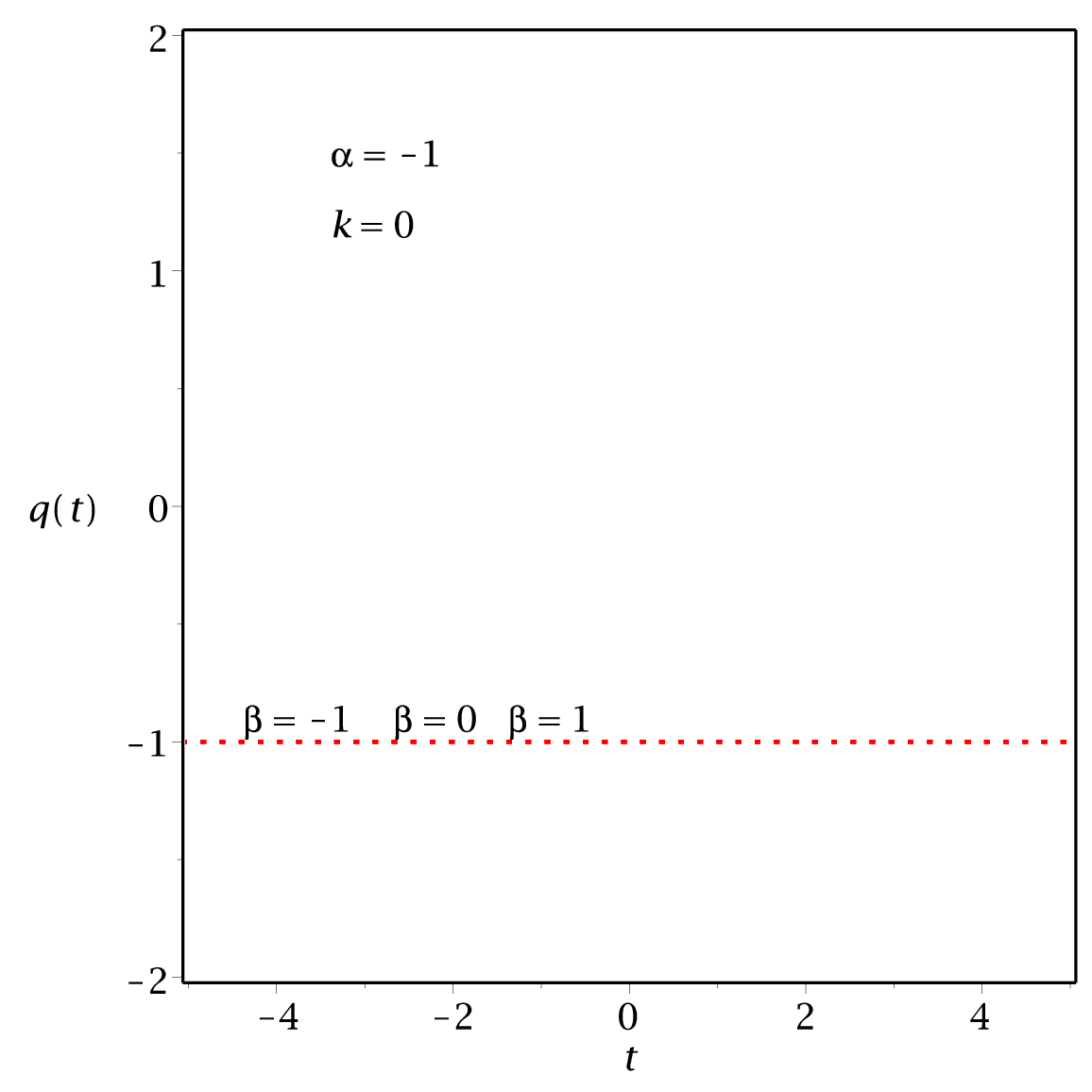}
\includegraphics[width=7cm]{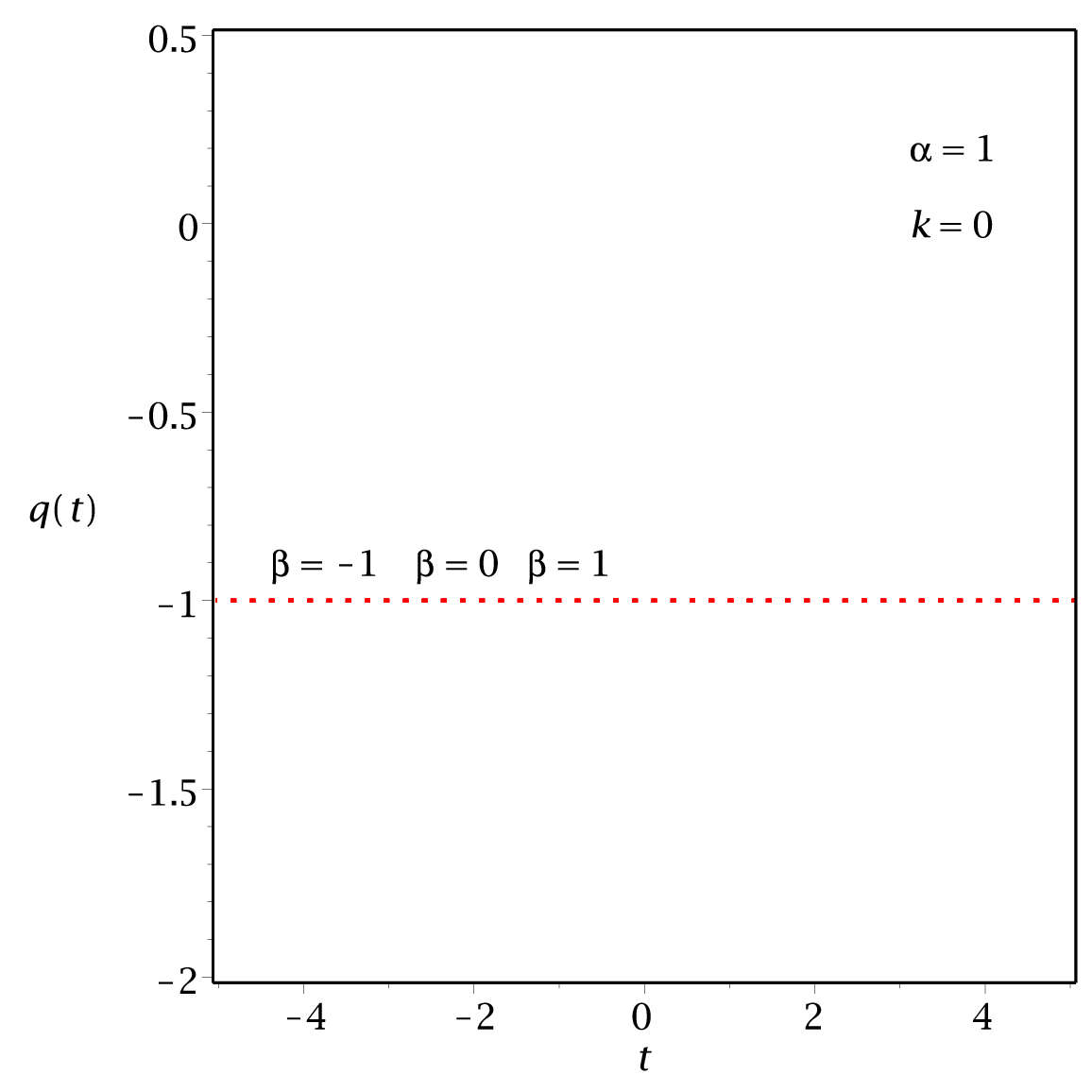}
\includegraphics[width=7cm]{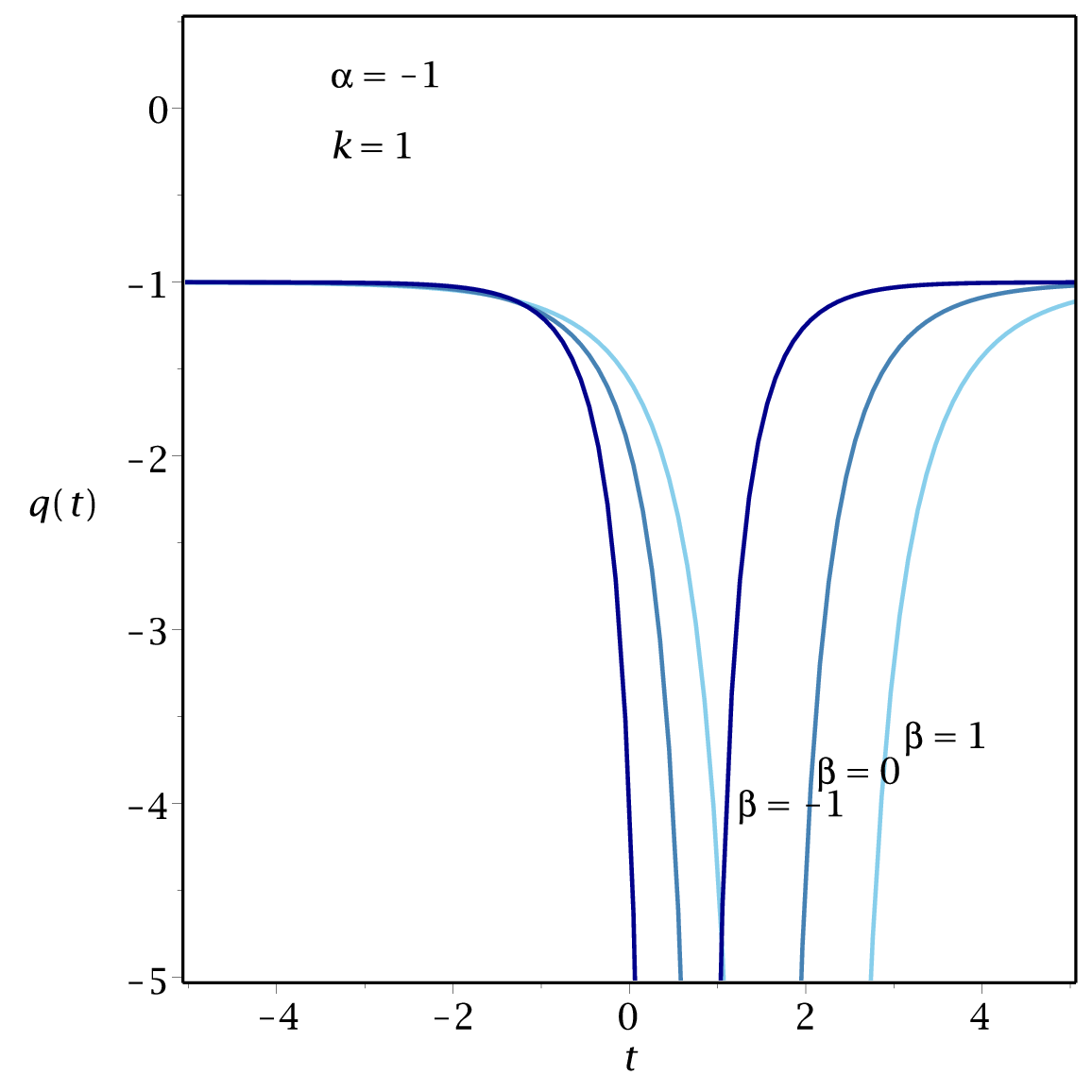}
\includegraphics[width=7cm]{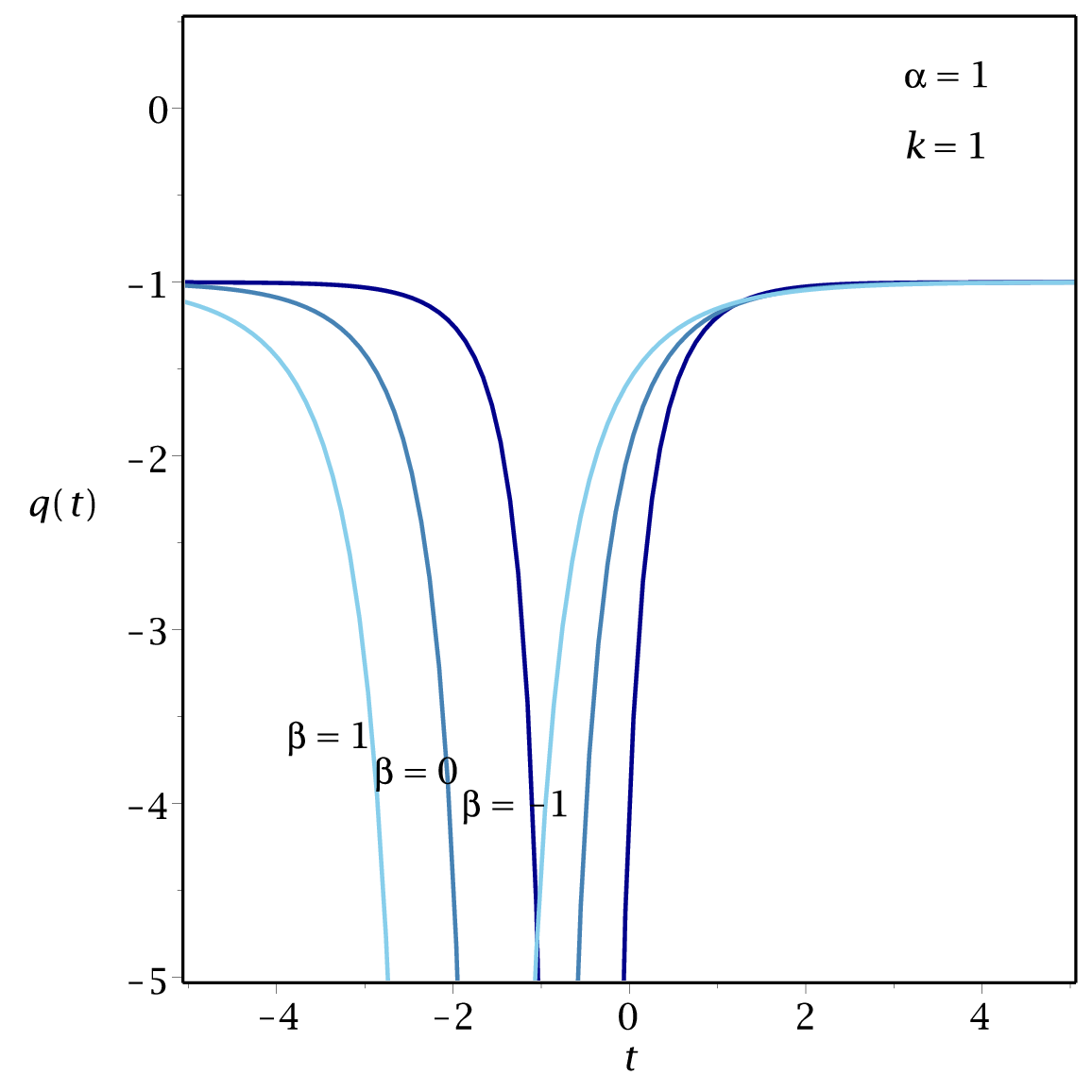}
\caption{Deceleration parameter evolution in time, derived for a fluid with constant energy density, equations (\ref{Bt1})-(\ref{Bt3}), for different values of $k=-1, 0, 1$, $\beta=-1,0,1$ and $\alpha= \pm 1.$}
\label{fig5}
\end{figure}

According to the solutions (\ref{Bt1})-(\ref{Bt3}), the system is always accelerated, as can be see in Figure \ref{fig5}. {Independently of the $k$ value, all the solutions converge asymptotically to $-1$ in $t \rightarrow \pm \infty$. The minimum acceleration occurs for $k=-1$ and $t \sim t_{sing}$, exactly at point where the scale factor change of signal, while for  $k=1$  the acceleration reaches the maximum divergent value $+\infty$, when the scale factor is zero, as show the Figure \ref{fig1}.}

\subsection{ {Dynamics of the vacuum solution with $\Lambda=0$}}

The vacuum solution given by equations (\ref{Bt0}) has its validity established within the parametric space where {$\beta >  -2$ and $k \le 0$ or $\beta < -2$ and $k \ge 0$}. { For the particular case $k=0$, the solution is static, while, for $k\neq 0$, $a(t)$ is linear and symmetric in time. According to (\ref{Ht}), the system expands in a constant rate.}

\section{Analysis of the energy conditions}

In this section, by analogy to what is usually done when we have the Einstein equations with cosmological constant, { in which  the term with $\Lambda$ is treated as a component of the energy momentum tensor, in the right side, let us consider the aether term} on the right side of the field equations and treat it as an extra component of the energy-momentum tensor. It is important to note that, as far as we know, there is no formulation for the energy conditions in the EA theory \cite{Garfinkle1}.

{Let us now, before the analysis of the energy conditions,
discuss in detail about the differences of the GR and EA theories, at least, for the strong energy condition. This condition is usually used
to define the existence of a dark energy fluid in the GR theory.
Let us recall that the weak, null and dominant energy conditions \cite{Hawking1973} are obtained imposing conditions of
physical reasonability on the matter fluid, i.e., for any observer, i) the density energy
must be positive; ii) the pressure must not exceed the energy density;
iii) the sound velocity in the fluid must not be greater than the vacuum
light speed. However, the strong energy condition ($\rho+3p\ge 0$, for an
isotropic fluid in the GR theory) comes from a geometrical condition on the tensor Riemann. The term $R_{ab} v^a v^b$ ($v^a$ is any timelike vector) in the analogous Raychaudhuri equation for timelike geodesics \cite{Hawking1973} must contribute negatively to ensure the {convergence of a}
congruence of these geodesics {, which defines the attractivity of the gravitation}. Thus, this condition must also be valid in the EA theory considering $T^{aether}_{ab}$ as part of an effective energy-momentum tensor $T^e_{ab}$, where
\bq
T^e_{ab}=T^{aether}_{ab}+8\pi G T^{matter}_{ab}.
\lb{Tabe}
\eq
We assume that the effective energy-momentum tensor corresponds to an isotropic fluid
with an energy density $\rho_e$, pressure $p_e$,  given by
\bq
T^{e}_{ab}=(\rho_e+p_e) w_a w_b + p_e g_{ab},
\lb{TGR}
\eq
and $w^a=\delta^a_t$ is a unit timelike vector representing the fluid velocity. We are also assuming a comoving reference.
Thus, following \cite{Hawking1973} we must have
\bq
R_{ab} w^a w^b \ge 0 \Rightarrow T^{e}_{ab} w^a w^b \ge \frac{1}{2} T^e w^a w^b \Rightarrow \rho_e +3p_e \ge 0.
\lb{Rab}
\eq
The components of the effective energy-momentum tensor are given by}
\bq
T^{e}_{tt}=T^{matter}_{tt}+T^{aether}_{tt}=8\pi G \rho-\frac{3\beta}{2} \frac{\dot a^2}{a^2}=\rho_e,
\lb{Te1}
\eq
\bqn
T^{e}_{rr}&=&T^{matter}_{rr}+T^{aether}_{rr}=\nb\\
&&=\frac{1}{1-k r^2}\left[8\pi G p a^2+\frac{\beta}{2}\left( \dot a^2+2 a \ddot a\right)\right]=p_e\frac{a^2}{1-k r^2},
\lb{Te2}
\eqn
\bq
T^{e}_{\theta\theta}=T^{matter}_{\theta\theta}+T^{aether}_{\theta\theta}=8\pi G p r^2 a^2+\frac{r^2 \beta}{2}  \left[\dot a^2+2 a \ddot a\right]=p_e r^2 a^2,
\eq
\bq
T^{e}_{\phi\phi}=T^{e}_{\theta\theta} \sin^2 \theta.
\eq

Using the equations (\ref{Te1}) and (\ref{Te2}) we get an equation of state {for the effective fluid}
\bq
p_e+\rho_e=8 \pi G(\rho+p) + \beta \frac{d}{dt}\left(\frac{\dot a}{a}\right).
\eq
{On the other hand, the strong energy condition becomes
{
\bq
\lb{SECef}
\rho_e+3p_e\ge 0 \Rightarrow \rho+3p \ge -\frac{3\beta}{8\pi G}\frac{\ddot a}{a}.
\eq
Using the field equation (\ref{Fried1}) we get $\rho+3p > 0$ if $\beta+2 >0$.
Note that in this model the strong energy condition for the effective energy momentum tensor is the same
of the matter energy momentum tensor.
}
In this equation we can see that if $\beta=0$ we recover the strong energy condition for the GR theory. In this case, a perfect fluid with equation of state given by $p=-\rho_0$, which represents a vacuum solution with $\Lambda > 0$, corresponds to a dark energy fluid. However, if $\beta \neq 0$ the strong energy condition is modified by the aether term and, therefore, $p=-\rho_0$ may not correspond a dark energy fluid. In order to verify this  we analyze how the aether term affects this energy condition.
}
From now on we will focus only on the contribution of the term from the aether. Our purpose is to find out what kind of fluid in GR theory would play a role analogous to it. Then, looking only at the term arising from the aether at the effective energy-momentum tensor, we have
\bq
T^{aether}_{tt}=-\frac{3\beta}{2} \frac{\dot a^2}{a^2}=\rho_{aether},
\lb{TEA1}
\eq
\bq
T^{aether}_{rr}=\frac{1}{1-k r^2}\left[\frac{\beta}{2}\left( \dot a^2+2 a \ddot a\right)\right]=p_{aether}\frac{a^2}{1-k r^2},
\lb{TEA2}
\eq
\bq
T^{aether}_{\theta\theta}=\frac{r^2 \beta}{2}  \left[\dot a^2+2 a \ddot a\right]=p_{aether} r^2 a^2,
\lb{TEA3}
\eq
\bq
T^{aether}_{\phi\phi}=T^{aether}_{\theta\theta} \sin^2 \theta.
\lb{TEA4}
\eq
Note that, in the case of perfect fluid considered here, all the aether terms depend on the energy density of the matter, $\rho_0$, as can be seen in (\ref{omega}). Bellow we will look at each solution separately, following the same classification used in Section 3.

\subsection{{Vacuum solutions with $\Lambda>0$}}

For solutions (\ref{Bt1})-(\ref{Bt3}), equations (\ref{TEA1})-(\ref{TEA3}) furnish,

for $k=-1$
\begin{equation}
\rho_{aether}=\frac{-8\rho_{{0}}{\it G}\,\pi \,\beta\, \left( 3(\beta+2)+{{\rm e}^{{\frac {-8\alpha\,\sqrt {3}\sqrt {\pi }\sqrt { \left( \beta+2 \right) {\it G}\,\rho_{{0}}}t}{3(\beta+2)}}}} \right) ^{2}} {\left( -3(\beta+2)+{{\rm e}^{{\frac {-8\alpha\,\sqrt {3}\sqrt {\pi }\sqrt {\left( \beta+2 \right) {\it G}\,\rho_{{0}}}t}{3(\beta+2)}}}} \right) ^{2} \left( \beta+2 \right)},
\end{equation}
\begin{eqnarray}
p_{aether}=& &-\frac{8\,\beta\,\pi \,{\it G}\,\rho_{{0}}}{\left( \beta+2 \right)\left( -3(\beta+2)+{{\rm e}^{-8/3\,{\frac {\alpha\,\sqrt {3}\sqrt {\pi }\sqrt { \left( \beta+2 \right) {\it G}\,\rho_{{0}}}t}{\beta+2}}}}\right) ^{2}} \nonumber\\
& &\times \left( -9(\beta+2)^{2}-{{\rm e}^{-16/3\,{\frac {\alpha\,\sqrt {3}
\sqrt {\pi }\sqrt { \left( \beta+2 \right) {\it G}\,\rho_{{0}}}t}{
\beta+2}}}}\right.\nonumber\\
& &\left. +2\,{{\rm e}^{-8/3\,{\frac {\alpha\,\sqrt {3}\sqrt {\pi }\sqrt { \left( \beta+2 \right) {\it G}\,\rho_{{0}}}t}{\beta+2}}}}\left(\beta+2 \right) \right).
\end{eqnarray}

{
and we have for the energy conditions,
\begin{equation}
\rho_{aether}+p_{aether}=-\frac{64\,\beta\,\pi \,{\it G}\,\rho_{{0}}{{\rm e}^{-8/3\,{
\frac {\alpha\,\sqrt {3}\sqrt {\pi }\sqrt { \left( \beta+2 \right) {
\it G}\,\rho_{{0}}}t}{\beta+2}}}} }{\left( -3(\beta+2)+{{\rm e}^{-8/3\,
{\frac {\alpha\,\sqrt {3}\sqrt {\pi }\sqrt { \left( \beta+2 \right) {
\it G}\,\rho_{{0}}}t}{\beta+2}}}} \right) ^{2}},
\end{equation}
\begin{eqnarray}
\rho_{aether}-p_{aether}=& & -\frac{16\rho_{{0}}{\it G}\,\pi \,\beta\,}{\left( \beta+2 \right) \left( -3(\beta+2)+{{\rm e}^{-8/3\,{\frac {\alpha\,\sqrt {3}\sqrt {\pi }\sqrt { \left( \beta+2 \right) {\it G}\,\rho_{{0}}}t}{\beta+2}}}} \right) ^{2}}\nonumber\\
& &\times \left(2\,(\beta+2)\,{{\rm e}^{-8/3\,{\frac {\alpha\,\sqrt {3}\sqrt {\pi }\sqrt { \left( 
\beta+2 \right) {\it G}\,\rho_{{0}}}t}{\beta+2}}}}\right.+\nonumber\\
& & \left.{{\rm e}^{-16/3\,{\frac {\alpha\,\sqrt {3}\sqrt {\pi }
\sqrt { \left( \beta+2 \right) {\it G}\,\rho_{{0}}}t}{\beta+2}}}}+9(\beta+2
)^2\right),
\end{eqnarray}
\begin{equation}
\rho_{aether}+3p_{aether}=16\,{\frac {\rho_{{0}}{\it G}\,\pi \,\beta\,}{\beta+2}},
\end{equation}
}

for $k=0$,
\begin{equation}
\rho_{aether}=-p_{aether}=-8\,{\frac {\rho_{{0}}{\it G}\,\pi \,\beta\,}{\beta+2}},
\end{equation}
{
and we have for the energy conditions,
\begin{equation}
\rho_{aether}+p_{aether}=0,
\end{equation}
\begin{equation}
\rho_{aether}-p_{aether}=-16\,{\frac {\rho_{{0}}{\it G}\,\pi \,\beta\,}{\beta+2}},
\end{equation}
\begin{equation}
\rho_{aether}+3p_{aether}=16\,{\frac {\rho_{{0}}{\it G}\,\pi \,\beta\,}{\beta+2}},
\end{equation}
}

for $k=1$,
\begin{equation}
\rho_{aether}=\frac{-8\rho_{{0}}{\it G}\,\pi \,\beta\, \left( -3(\beta+2)+{{\rm e}^{{\frac {-8\alpha\,\sqrt {3}\sqrt {\pi }\sqrt { \left( \beta+2 \right) {\it G}\,\rho_{{0}}}t}{3(\beta+2)}}}} \right) ^{2}} {\left( 3(\beta+2)+{{\rm e}^{{\frac {-8\alpha\,\sqrt {3}\sqrt {\pi }\sqrt {\left( \beta+2 \right) {\it G}\,\rho_{{0}}}t}{3(\beta+2)}}}} \right) ^{2} \left( \beta+2 \right)},
\end{equation}
\begin{eqnarray}
p_{aether}=& &\frac{8\,\beta\,\pi \,{\it G}\,\rho_{{0}}}{\left( \beta+2 \right)\left( 3(\beta+2)+{{\rm e}^{-8/3\,{\frac {\alpha\,\sqrt {3}\sqrt {\pi }\sqrt { \left( \beta+2 \right) {\it G}\,\rho_{{0}}}t}{\beta+2}}}} \right) ^{2}} \nonumber\\
& &\times \left( 9(\beta+2)^2+{{\rm e}^{-16/3\,{\frac {\alpha\,\sqrt {3}
\sqrt {\pi }\sqrt { \left( \beta+2 \right) {\it G}\,\rho_{{0}}}t}{
\beta+2}}}}\right.\nonumber \\
& & \left.+2\,{{\rm e}^{-8/3\,{\frac {\alpha\,\sqrt {3}\sqrt {
\pi }\sqrt { \left( \beta+2 \right) {\it G}\,\rho_{{0}}}t}{\beta+2}}}
} \left( \beta+2 \right)  \right),
\end{eqnarray}
{
and we have for the energy conditions,
\begin{equation}
\rho_{aether}+p_{aether}=\frac{64\,\beta\,\pi \,{\it G}\,\rho_{{0}}{{\rm e}^{-8/3\,{
\frac {\alpha\,\sqrt {3}\sqrt {\pi }\sqrt { \left( \beta+2 \right) {
\it G}\,\rho_{{0}}}t}{\beta+2}}}} }{\left( 3(\beta+2)+{{\rm e}^{-8/3\,
{\frac {\alpha\,\sqrt {3}\sqrt {\pi }\sqrt { \left( \beta+2 \right) {
\it G}\,\rho_{{0}}}t}{\beta+2}}}} \right) ^{2}},
\end{equation}
\begin{eqnarray}
\rho_{aether}-p_{aether}=& & -\frac{16\rho_{{0}}{\it G}\,\pi \,\beta\,}{\left( \beta+2 \right)\left( 3(\beta+2)+{{\rm e}^{-8/3\,{\frac {\alpha\,\sqrt {3}\sqrt {\pi }\sqrt { \left( \beta+2 \right) {\it G}\,\rho_{{0}}}t}{\beta+2}}}}\right) ^{2}}\nonumber\\
& &\times\left(-2(\beta+2)\,{{\rm e}^{-8/3\,{\frac {\alpha\,\sqrt {3}\sqrt {\pi }\sqrt { \left( 
\beta+2 \right) {\it G}\,\rho_{{0}}}t}{\beta+2}}}}\right.+\nonumber\\
& & \left.{{\rm e}^{-16/3\,{\frac {\alpha\,\sqrt {3}\sqrt {\pi }
\sqrt { \left( \beta+2 \right) {\it G}\,\rho_{{0}}}t}{\beta+2}}}}+9(\beta+2)^2\right), \nonumber\\
\end{eqnarray}
\begin{equation}
\rho_{aether}+3p_{aether}=16\,{\frac {\rho_{{0}}{\it G}\,\pi \,\beta\,}{\beta+2}}.
\end{equation}
}

\hskip2em Note that we must choose $\beta\leq 0$ in order to ensure a non-negative energy density, since $\beta\geq-2$, that is, the parameter $\beta$ should be restricted to the interval $-2 <\beta\leq 0$. Let us analyze the results for each of the values chosen for $ k $. 

\hskip2em In the case  $k=-1$ it is easy to see that the energy conditions are satisfied except the latter, which means that the aether plays a role equivalent to a dark energy fluid in the GR theory {in the sense that $p_{aether}$ is sufficiently negative in order to violate the strong energy condition, although $p_{aether}\ne -\rho_{aether}$} \cite{Chan}. {Thus, we have the matter fluid with $p=-\rho_0$ and the aether component acting as a dark energy fluid.}

\hskip2em In a similar but not identical way, for $k=0$, we have only the last energy condition violated and we can interpret the aether's component as a dark energy fluid  playing an analogous role placed by the positive cosmological constant in GR theory, since $p_{aether}=-\rho_{aether}$, with $\rho_{aether}=constant$, as can be seen above. {Then, we have the matter fluid with $p=-\rho_0$ and the aether component acting as a dark energy fluid similarly to an "aether cosmological constant" ($\Lambda_{aether}$), since $p_{aether}=-\rho_{aether}$.}

\hskip2em On the other hand, for $ k = 1 $, in addition to the last energy condition, the former is also violated, implying that aether behaves equivalently to an phantom fluid in the GR theory \cite{Chan}. {Again, we have the matter fluid with $p=-\rho_0$ but the aether component acting as a phantom energy fluid, since it violates both strong and null energy conditions.}

\hskip2em Accordingly, we can concludes that, for the model described here, the aether's component reinforces the accelerated behavior of the fluid.

\subsection{ {Vacuum solution with $\Lambda=0$}}

\begin{equation}
\rho_{aether}=-\frac{3}{2}\beta/t^2,
\end{equation}
\begin{equation}
p_{aether}=\frac{1}{2}\beta/t^2.
\end{equation}
Here again we must have $\beta\leq 0$ in order to ensure a non-negative energy density. In addition with the {conditions imposed by equation (\ref{Bt0}) we must have  $-2<\beta\leq 0$ ($k\le 0$) or $\beta < -2$ ($k\geq 0$).} The energy conditions are given by
\begin{equation}
\rho_{aether}+p_{aether}=-\beta/t^2,
\end{equation}
\begin{equation}
\rho_{aether}-p_{aether}=-2\beta/t^2,
\end{equation}
\begin{equation}
\rho_{aether}+3p_{aether}=0.
\end{equation}
Therefore, {in this case where there is no matter} the aether's component satisfy all the energy conditions  and both  $\rho_{aether}$ and $p_{aether}$ decreases over time. In the context of the GR theory it could be interpreted as a non-accelerated expansion cosmological solution.

\section{Conclusions}

In the present work we have analyzed the possible solutions allowed by the theory of EA which breaks the Lorentz invariance. Considering a  Friedmann-Lema{\^{\i}}tre-Robertson-Walker (FLRW) metric, we have obtained exact solutions for two particular cases: (i) a perfect fluid with constant energy density
{($\Lambda>0$)}, and (ii) a fluid with {zero} energy density
{($\Lambda=0$)}. 
Our solutions show
that the EA and GR theories do not differentiate to each other only by the
coupling constant. This difference is clearly shown because of the existence
of singularities that there are not in GR theory. This characteristic appears in the solutions with
{$\Lambda>0$ and $k=1$} as well as with {$\Lambda=0$}, where this last one depends only on the aether field.  Besides, we consider {the effective energy momentum tensor} in the Raychaudhuri equation and discuss the meaning of the strong energy condition in this scenario. {It is important to notice that the de Sitter solution are obtained from our solutions only imposing  $c_1+3c_2+c_3=0$ ($\beta=0$) and $c_1=-c_4$ ($G=G_N$), simultaneously.
The solutions admit an expanding or contracting system. A bounce, a  singular, a constant and an accelerated expansion { or contraction} solution{s} were also obtained.
 
{For ({$\Lambda>0$)} the Kretschmann scalar has singularities at $t_{sing} = -\frac{1}{2\alpha w}\ln(6+3\beta)$ only for $k=-1$ {($\beta \ne 0$ and $\alpha=\pm 1$)} where  $w=4\sqrt{\frac{\pi G \rho_0}{3(\beta+2)}}$.
{For the particular case $\beta=0$ this scalar is finite and constant, as expected in GR theory. The presence of singularities for
$\beta \ne 0$ shows a behavior that is completely different as predicted
in the GR theory.}

In the particular case of zero density energy {($\Lambda=0$)}, we have obtained a vacuum solution for the spacetime, for which the  Kretschmann scalar is not null but singular at the initial time, unless $\beta=0$. {As in the case where {$\Lambda>0$} and $k=-1$ this result also
highlights another distinction between the GR and EA theories.}

The analysis of energy conditions {reveals that in the
hypothesis of { $\Lambda>0$}, we have an aether dark
energy fluid with $p_{aether} \ne -\rho_{aether}$ (for $k=-1$); an aether dark
energy fluid with $p_{aether} = -\rho_{aether}$ where $\rho_{aether}$ is constant (for $k=0$); an aether  phantom energy fluid with $p_{aether} \ne -\rho_{aether}$ (for $k=1$). In the hypothesis
that {$\Lambda=0$}, the aether fluid behaves as a normal energy fluid.
}

{Therefore the results} show that the presence of aether contributes significantly for the accelerated expansion of the system for {$\Lambda>0$}. On the other hand, for the vacuum case {with $\Lambda=0$}, the energy conditions are all satisfied.

\section*{Acknowledgments}

The financial assistance from FAPERJ/UERJ (MFAdaS) is gratefully acknowledged. The author (RC) acknowledges the financial support from FAPERJ (no.E-26/171.754/2000, E-26/171.533/2002 and E-26/170.951/2006). MFAdaS and RC also acknowledge the financial support from Conselho Nacional de Desenvolvimento Cient\'{\i}fico e Tecnol\'ogico - CNPq - Brazil.  The author (MFAdaS) also acknowledges the financial support from Financiadora de Estudos e Projetos - FINEP - Brazil (Ref. 2399/03). VHS gratefully acknowledges the financial support from FAPERJ though Programa P\'{o}s-doutorado Nota 10. VHS also thanks Jailson Alcaniz for the hospitality at Observat\'{o}rio Nacional (ON), and Jos\'{e} Abdalla Helay\"{a}l-Neto and Sofiane Faci at Centro Brasileiro de Pesquisas F\'{i}sicas (CBPF). (MC) thanks ON and UFRJ for their hospitality.
{We would like to thank the anonymous referee for the
valuable suggestions and questions that improved this work.}


\section*{References}

\begin{thebibliography}{88}

\bibitem{Jacobson:2000xp} 
T.~Jacobson and D.~Mattingly,
Phys.\ Rev.\ D {\bf 64}, 024028 (2001)
[arXiv:0007031]

\bibitem{Elliott:2005va} 
J.~W.~Elliott, G.~D.~Moore and H.~Stoica,
JHEP {\bf 0508}, 066 (2005)
[hep-ph/0505211]

\bibitem{Eling:2003rd} 
C.~Eling and T.~Jacobson,
Phys.\ Rev.\ D {\bf 69}, 064005 (2004)
[arXiv:0310044]

\bibitem{GrEAsser:2005bg} 
M.~L.~GrEAsser, A.~Jenkins and M.~B.~Wise,
Phys.\ Lett.\ B {\bf 613}, 5 (2005)
[hep-th/0501223]

\bibitem{Foster:2007gr} 
B.~Z.~Foster,
Phys.\ Rev.\ D {\bf 76}, 084033 (2007)
[arXiv:0706.0704]

\bibitem{Yagi:2013ava} 
K.~Yagi, D.~Blas, E.~Barausse and N.~Yunes,
Phys.\ Rev.\ D {\bf 89}, no. 8, 084067 (2014)
Erratum: [Phys.\ Rev.\ D {\bf 90}, no. 6, 069902 (2014)]
Erratum: [Phys.\ Rev.\ D {\bf 90}, no. 6, 069901 (2014)]
[arXiv:1311.7144]

\bibitem{Gong:2018cgj} 
Y.~Gong, S.~Hou, D.~Liang and E.~Papantonopoulos,
Phys.\ Rev.\ D {\bf 97}, 084040 (2018)
[arXiv:1801.03382]

\bibitem{Oost:2018tcv} Oost, J. Mukohyama, S., Wang, A., Phys. Rev. D {\bf 97}, 124023 (2018)

\bibitem{Foster:2005dk} 
B.~Z.~Foster and T.~Jacobson,
Phys.\ Rev.\ D {\bf 73}, 064015 (2006)
[arXiv:0509083]

\bibitem{Barrow2011} Barrow, J.D., Phys. Rev. D {\bf 85}, 047503 (2012)

\bibitem{Solomon2014} Solomon, A.R. and Barrow, J.D., Phys. Rev. D {\bf 89}, 024001 (2014)

\bibitem{Dina} Ding, C., Wang, A., Wang, X., Phys. Rev. D {\bf 92}, 084055 (2015)

\bibitem{Dinb} Ding, C., Wang, A., Wang, X., Zhu, T., Phys Rev D {\bf 94} 124034 (2016) 

\bibitem{Ho} Ho, F.H. Zhang, S.J. Liu, H.S. Wang, A., Phys. Let. B {\bf 782}, 723 (2018)

\bibitem{Oostb} Oost, J., Bhattacharjee, M, Wang, A.,  
[arXiv:1804.01124]

\bibitem{Bhattacharjee} Bhattacharjee, M. Mukohyama, S., Wan, M.B., Wang, A., Phys. Rev. D {\bf 98}, 064010 (2018) 

\bibitem{Carroll:2004ai} 
S.~M.~Carroll and E.~A.~Lim,
Phys.\ Rev.\ D {\bf 70}, 123525 (2004)
[hep-th/0407149]

\bibitem{Garfinkle} D. Garfinkle,  C. Eling and T. Jacobson,  Phys. Rev. D 76, 024003 (2007)

\bibitem{Zlosnik_2007} T.G. Zlosnik, P.G. Ferreira and G.D. Starkman, 
Phys. Rev. D 75, 044017 (2007)

\bibitem{Zlosnik_2008} T.G. Zlosnik, P.G. Ferreira and G.D. Starkman, 
Phys. Rev. D 77, 084010 (2008)

\bibitem{Zuntz_2010} J. Zuntz, T.G. Zlosnik, F. Bourliot, P.G. Ferreira, and
G.D. Starkman, 
Phys. Rev. D 81, 104015 (2010).

\bibitem{Chan} R. Chan, M. F. A. da Silva and J. F. Villas da Rocha, Mod. Phys. Lett. A, 24, 1137 (2009)

\bibitem{Garfinkle1} D. Garfinkle and T. Jacobson, Phys. Rev. Lett. 107, 191102 (2011)

{\bibitem{Jacobson08} T. Jacobson, (2008) [arxiv:0801.1547]}

{
{\bibitem{Mattingly2001} D. Mattingly and T. Jacobson, (2001) [arxiv:0112012]}
}

\bibitem{OS_1939} J. R.  Oppenheimer and H. Snyder, Phys. Rev. 56, 455 (1939)

\bibitem{Kolassis1988} C.A. Kolassis, N.O. Santos and D. Tsoubelis, 
    Astrophys. J. 327, 755 (1988)
    
\bibitem{Chan1989} R. Chan, J.P.S. Lemos, N.O. Santos and J.A. de F. Pacheco, Astrophys. J. 342, 976 (1989)

\bibitem{Sharif2010} M. Sharif and G. Abbas, Astrophys. Spa. Sci. 327, 285 (2010)

\bibitem{Herrera} L. Herrera, J. Jim�nez, and G. J. Ruggeri
Phys. Rev. D 22, 2305 (1980)

\bibitem{Stephani1990} H. Stephani, Cambridge University Press, {General Relativity: An Introduction to the Theory of Gravitational Field}, p. 268 (1990)

\bibitem{Rindler2006} W. Rindler, Oxford University Press, {Relativity: Special, General, and Cosmological}, p. 398 (2006)

\bibitem{Padmanabhan2010} T. Padmanabhan, Cambridge University Press, {Gravitation: Foundations and Frontiers}, p. 479 (2010)

\bibitem{d'Inverno1992} R. d'Inverno, Oxford University Press, {Introducing Einstein's Relativity}, p. 340 (1992)

\bibitem{Hawking1973} S.W. Hawking and G.F.R. Ellis, The Large Scale Structure of Space-time, Cambridge University Press, p. 124 (1973)

\bibitem{Milne} A. Benoit-Lévy, and G. Chardin, Astron. and Astrophys., 537, idA478 (2012) 

\end{thebibliography}
\end{document}